\documentclass{aa}
\usepackage{graphicx}
\usepackage{txfonts}
\usepackage{natbib} 

\pdfoutput=1
\def\ltsim{\hbox{\raise 2pt \hbox {$<$} \kern-1.1em \lower 4pt \hbox {$\sim$}}}
\def\ltapprox{\hbox{\raise 2pt \hbox {$<$} \kern-1.1em \lower 5pt \hbox
{$\approx$}}}
\def\gtsim{\hbox{\raise 2pt \hbox {$>$} \kern-1.1em \lower 4pt \hbox {$\sim$}}}
\def\gtapprox{\hbox{\raise 2pt \hbox {$>$} \kern-1.1em \lower 5pt \hbox
{$\approx$}}}

\def\eps{\varepsilon}

\def\arcmin{$^{\prime}$}

\begin{document}
   \title{Low-frequency study of two Giant Radio Galaxies:\\ 3C35 and 3C223}

   \subtitle{}

   \author{E. Orr{\`u} \inst{1,2}
           \and
           M. Murgia \inst{3,4}
           \and 
           L. Feretti \inst{4} 
           \and
           F. Govoni \inst{3} 
           \and
           G. Giovannini \inst{4,5} 
	   \and
           W. Lane \inst{6} 
           \and
           N. Kassim \inst{6} 
           \and 
           R. Paladino \inst{1}
          }

   \offprints{E. Orr{\`u} e.orru@astro.ru.nl}

   \institute{Institute of Astro- and Particle Physics, University of
     Innsbruck, Technikerstrasse 25/8,     A-6020, Innsbruck, Austria 
     \and
     Radboud University Nijmegen, Heijendaalseweg 135,   6525 AJ Nijmegen, The Netherlands
     \and
     INAF\,-\,Osservatorio Astronomico di Cagliari, Loc. Poggio dei Pini, Strada 54,
     I-09012 Capoterra (CA), Italy
     \and
     INAF\,-\,Istituto di Radioastronomia, Via Gobetti 101, I-40129 Bologna, Italy
     \and
     Dipartimento di Astronomia Universit{\`a} degli Studi di Bologna, 
     Via Ranzani 1, 40127 Bologna, Italy
     \and
     Naval Research Laboratory, Code 7213, Washington DC 20375-5320, USA
     }

   \date{Received; accepted}

 \abstract
   {}
   {Radio galaxies with a projected
linear size $\gtrsim$\,1 Mpc are classified as giant radio sources.
According to the current interpretation these are old sources which have evolved in a 
low-density ambient medium.
Since radiative losses are negligible at low frequency, extending spectral 
ageing studies in this frequency range will allow to determine the 
zero-age electron spectrum
injected and then to improve the estimate of the synchrotron age of the source.
}
   {We present Very Large Array images at 74 MHz and 327 MHz of two 
giant radio sources: 3C35 and 3C223. 
We performed a spectral study using 74, 327, 608 and 1400 GHz images. The spectral shape is estimated in different positions along the source. 
}
   {The radio spectrum follows a power-law in the hot-spots, while in
the inner region of the lobe the shape of the spectrum shows a
curvature at high frequencies.  This steepening is in agreement with
synchrotron aging of the emitting relativistic electrons.  In order to
estimate the synchrotron age of the sources, the spectra have been
fitted with a synchrotron model of emission. Using the models, we find
that 3C35 is an old source of 143$\pm$20\,Myr, while 3C223 is a younger source of 72$\pm$4\,Myr.}
   {}

   \keywords{Instrumentation:interferometers - Techniques: interferometric - Astroparticle physics - Galaxies: active - Radio continuum: galaxies - Radiation mechanisms: non - thermal}

\maketitle
%

\section{Introduction}

\begin{table*}
\caption{Summary of radio observations and images.
Col. 1: source name;
Col. 2, 3: radio pointing position;
Col. 4: observing frequency;
Col. 5: bandwidth;
Col. 6: date of observations;
Col. 7: time of integration;
Col. 8: VLA configuration;
Col. 9: resolution;
Col. 10: Position Angle;
Col. 11:rms.}
\label{obs}
\begin{tabular}{ccccccccccc}
\noalign{\smallskip}
\hline
\noalign{\smallskip}
Name   &$\alpha$(J2000) & $\delta$(J2000)  &${\nu}$& $\Delta\nu$ & Date & Duration &Array& HPBW&PA&rms\\
      & ($h$ $m$ $s$)  & (\degr\ \arcmin\ \arcsec)& MHz& MHz&  &hours&&\arcsec$\times$\arcsec&\degr&mJy/beam\\
\hline
3C35 & 01 12 02.20 & +49 28 35.00 & 73.8 & 1.562 &23-11-2003&6&B&93$\times$64&-85&95 \\
& & &327.5&3.125&23-11-2003&6&B&23$\times$17&-85&1.3 \\
& & &327.5-321.5&3.125&21-03-2004&3.5&C&55$\times$50&15&2.3\\
& & &327.4&3.125&&9.5&B+C&27$\times$21&-88&1.0\\

\hline
3C223 &09 39 52.74&+35 53 58.20& 73.8 & 1.562 &16-11-2004&5&A&25$\times$24&-58&43\\
&&& 73.8 & 1.562 &03-03-2005&5&B&83$\times$73&-62&98\\
&&&73.8& 1.562 &&10&A+B&26$\times$25&-61&40\\
&&&327.3&6.25&16-12-2004&5&A&6$\times$5&-77&0.7\\
&&&328.9&6.25&03-03-2005&5&B&19$\times$16&-82&1.3\\
&&&327.3&6.25&&10&A+B&7$\times$6&-78&0.6\\
\hline
\noalign{\smallskip}
\end{tabular}
\end{table*}

\begin{table*}
\caption{Sources properties.}
\begin{center}
\begin{tabular}{cccccccc}
\noalign{\smallskip}
\hline                  
\noalign{\smallskip}    
Name   & $\alpha$(J2000) & $\delta$(J2000) & z & kpc$\slash\arcsec$&LAS& LLS & L$_{178\,MHz}$   \\
       & ($h$ $m$ $s$)  & (\degr\ \arcmin\ \arcsec) &  &&\arcmin &kpc&  W\,Hz$^{-1}$              \\               
\hline
3C35 &01 12 02.23 &$+$49 28 35.2 &  0.0673 &1.273&12.5&950& 10$^{26.09}$\\
3C223  &09 39 52.74  &$+$35 53 58.2  &0.1368 &2.393&5.4 &780&10$^{26.89}$\\
\hline
\noalign{\smallskip}
\label{source}
\end{tabular}
\end{center}
\begin{list}{}{}
\item[] 
Col. 1: source name; 
Col. 2: and 3: source coordinates from NASA/IPAC extragalactic database (NED) ; 
Col. 4: redshift by \cite{Burbidge1972} and \cite{SDSS}; 
Col. 5: arcsec to kpc conversion;
Col. 6: largest angular size;
Col. 7: largest linear size;
Col. 8: radio luminosity at 178 MHz \cite[]{Laing1980}.
\end{list}
\end{table*}

\begin{figure*}[th]
\begin{center}
\includegraphics[width=14cm]{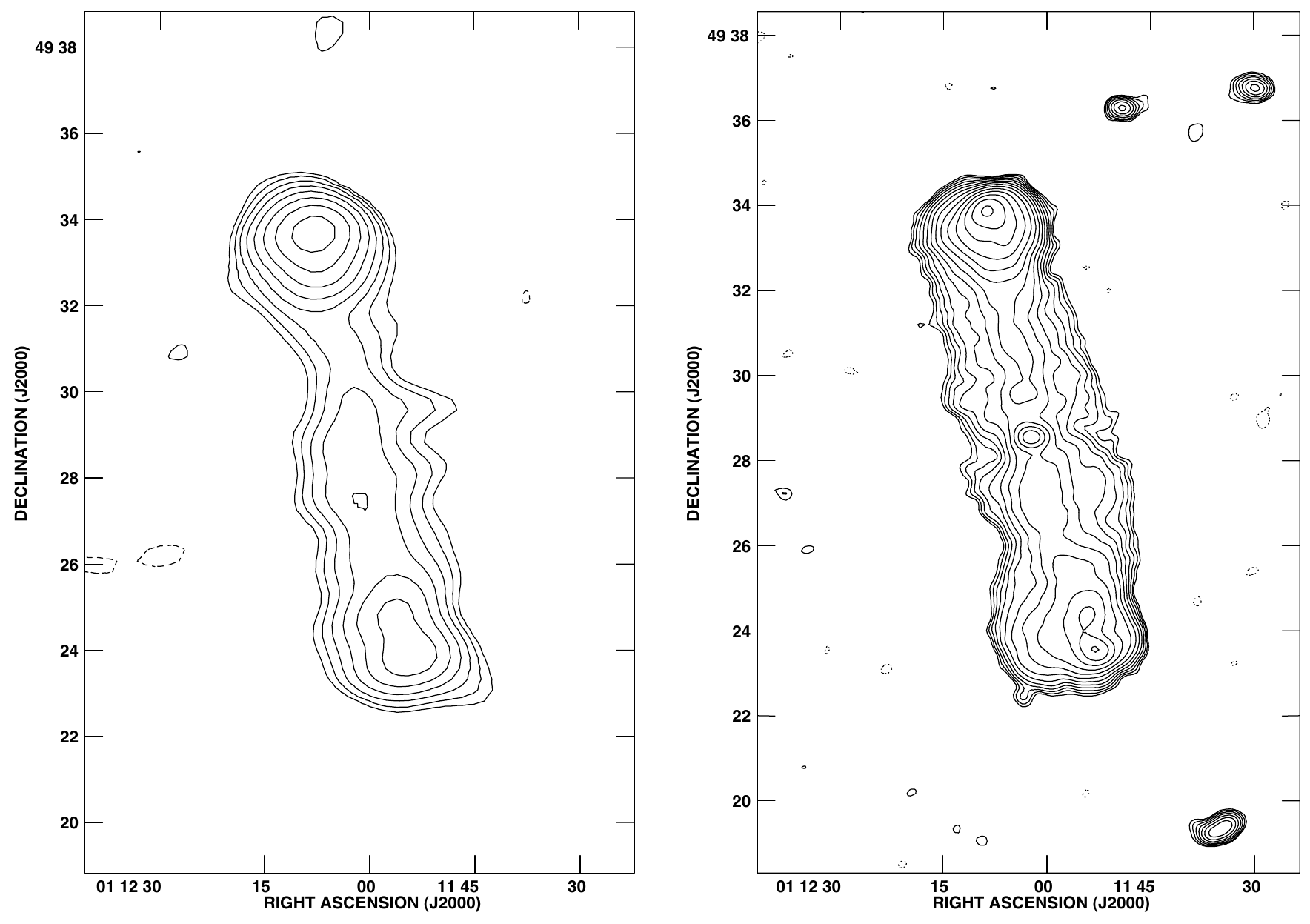}
\end{center}
\caption[]{Radio images of 3C35, all contours start at (3$\sigma$) and
  are scaled by $\sqrt{2}$. {\it Left}: VLA image at 74 MHz; the resolution is 93\arcsec$\times$64\arcsec\, with a PA=$-$85\degr, and the first two levels are at $-$285 and 285\,mJy/beam.
{\it Right}: 327 MHz VLA image. The image is obtained by combining the B and C configuration data, and the resolution is 27\arcsec$\times$21\arcsec\, with a PA=$-$88\degr. The first two levels of contours are $-$3 and 3\,mJy/beam.}
\label{3c35cntrnat}
\end{figure*} 

\begin{figure*}[th]
\begin{center}
\includegraphics[width=14cm]{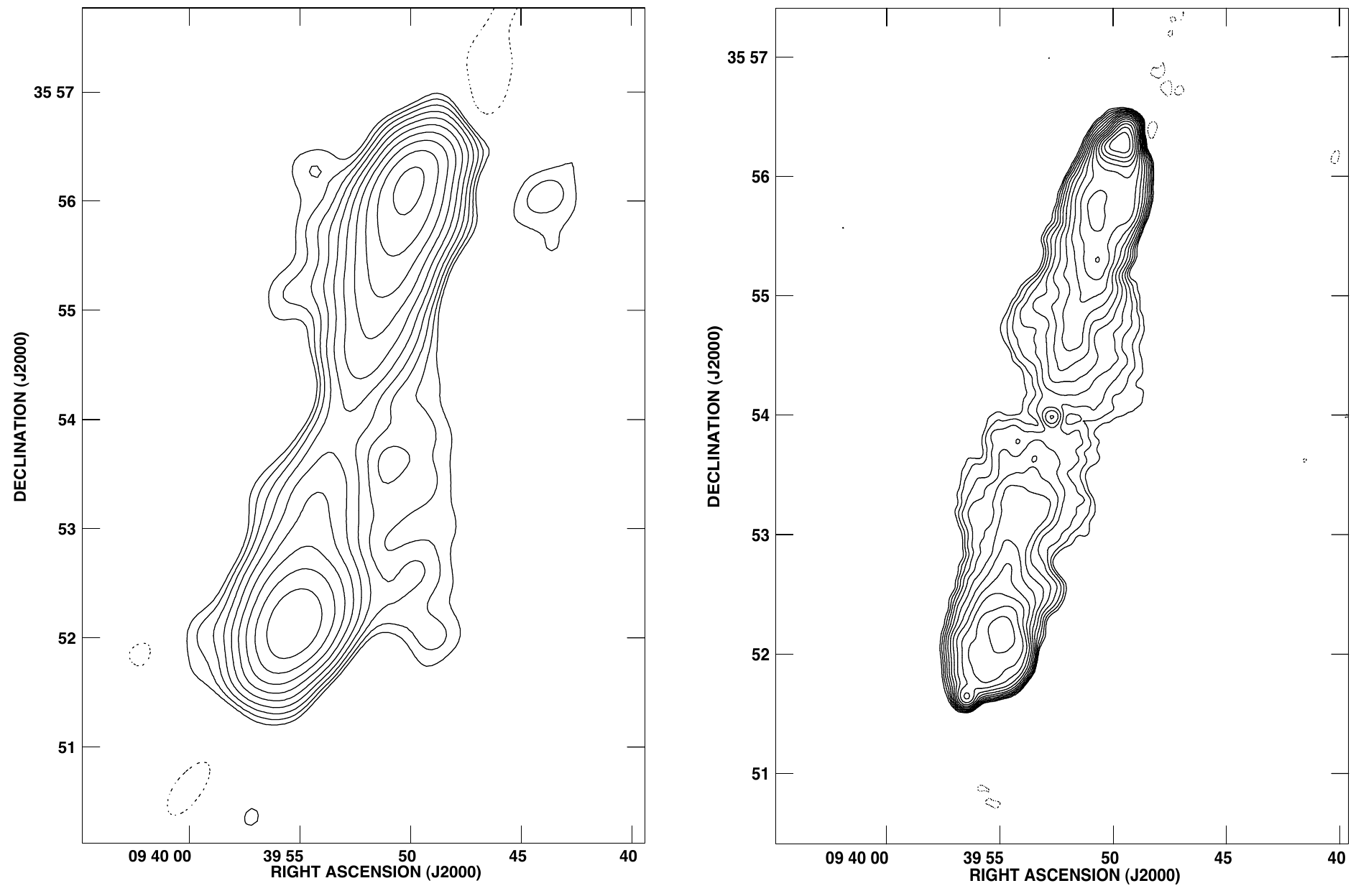}
\end{center}
\caption[]{Radio images of 3C223, all contours start at
  (3$\sigma$) and are scaled by $\sqrt{2}$. {\it Left}: VLA
  image at 74 MHz. The image is obtained by combining the data of A and B configurations;  
the resolution is 26\arcsec$\times$25\arcsec\, with a PA=$-$61\degr. The first two levels of contours are $-$120 and 120\,mJy/beam.
{\it Right}: 327 MHz VLA image, obtained by combining the data 
of A and B configurations; the resolution is 7\arcsec$\times$6\arcsec\, with a PA=$-$78\degr.  The first two levels of contours are $-$1.8 and 1.8\,mJy/beam.}
\label{3c223cntrnat}
\end{figure*} 

According to the standard model of active galactic nuclei (AGNs), at
the center of active galaxies resides a super-massive black hole. The
AGN is powered by an accretion disk surrounded by a torus of gas and
dust \cite[]{Blandford&Rees1974}.  The powerful radio emission
observed in classical double radio sources is produced by a bipolar
pair of jets; relativistic outflows of matter which originate in the
AGN.  They first propagate in the interstellar medium (ISM) and then in the
intergalactic medium (IGM) for a typical time of 10$^8$~yr
\cite[]{Scheuer1974} . The hot-spots are the regions where the energy
carried by the jets is diffused into the radio lobes.  The observed
diffuse radio emission is produced in the 'cocoon' or lobe, which is
formed by the built up jet material and/or energy in the region between
the core and the hot-spots.  The energy evolution of the cocoon can be
traced by observations and spectral studies of the radio lobes.  Radio
lobes expand, and, assuming the source is in the equipartition regime, the
pressure of the relativistic plasma in the lobe equals the pressure of
the external environment \cite[]{Begelman1984}.

The radio spectrum of radio galaxies is initially described by a
power-law. The final shape of the spectrum moves away from the
power-law showing a steepening at higher frequencies. This is due to
the competition between processes of energy injection and losses due
to adiabatic expansion, synchrotron emission and inverse Compton
scattering with the CMB photons, \citep{K,Kellermann1964,P}.  The
initial models developed to interpret these spectra assumed a uniform
and constant magnetic field and an isotropic injection of electrons
\citep[e.g.,][hereafter KP and JP]{K,P,JP}.  If the former assumptions
are satisfied it is possible, in theory, to use the synchrotron
spectrum to estimate the age of the radiating particles.

Many authors \citep[e.g.,][]{Tribble1993,Eilek1997,Blundell2000} argue
that the observed filamentary structures in the radio lobes
\citep[e.g. for Cygnus A,][]{Carilli1991} can be interpreted as the
effect of inhomogeneous magnetic fields on the synchrotron
emission. However, \cite{Kaiser2000} demonstrated that the spatial
distribution of the synchrotron radio emission can be used to estimate
the age for FRII sources \cite[]{Fanaroff1974}. Furthermore, based on
the dynamical and radiative self-similar models in
\cite{Kaiser&Alexander1997} and \cite{KaiserThorpe1997},
\cite{Kaiser2000} developed a 3-dimensional model of the synchrotron
emissivity of the cocoons of powerful FRII radio sources. The
projection along the line of sight (LOS) of the 3D model can be easily
compared with radio observations.

X-ray emission related to the lobes has been detected in a number of
radio sources.  This is attributed to the inverse Compton (IC)
scattering of microwave background photons. The direct estimates of
the magnetic field ($B_{IC}$) obtained from the combination of the
X-ray IC flux and the radio synchrotron spectrum, give values near to
those found with the equipartition ($B_{eq}$) assumption
\citep[e.g.,][]{Croston2004,Croston2005}. Moreover, the above-mentioned comparisons suggest that lobes are not overpressured at the late
stages of evolution of radio galaxies  \citep{Croston2004,Croston2005,Konar2009}. On the other hand, a strong
variation of the X-ray/radio flux ratio across the lobes has been
found \citep{Isobe2002,Hardcastle2005,Goodger2008}. This cannot be
explained with models in which either the electron energy spectrum or
the magnetic field vary independently as function of position in the
lobes, but it is consistent with models in which both vary together as
function of position.

\noindent
Among radio galaxies (RG), those with a projected linear size
$\gtsim$~1\,Mpc\footnote{Throughout we adopt
H$_0$\,=\,71\,km\,s$^{-1}$\,Mpc$^{-1}$, $\Omega_m$\,=\,0.27,
$\Omega_{\Lambda}$\,=\,0.73 \cite[]{Spergel2003}. Many radio galaxies
have been classified as giant in the past using a different set of
cosmological parameters. For this reason some GRG could have a linear
size slightly less than 1Mpc.} are defined as giant radio galaxies
(GRG).  In the complete sample of 3CR radio sources \cite[]{Laing1983}
around 6$\%$ of the sources are giants; there are about 100 known. GRG
typically have radio powers below $10^{26.5}$ W~Hz$^{-1}$~sr$^{-1}$,
have linear sizes less than 3 Mpc, and are observed at redshifts
$z<$~0.25, even though $z<$~0.5 could be assumed as an upper limit
\citep{Ishwara-Chandra1999,Schoenmakers2000,Lara2004,Saripalli2005,Machalski2007}. The
P-D diagram \citep{Lara2004,Ishwara-Chandra1999} shows a dearth of
high luminosity GRG, as predicted by evolutionary models
\citep{Blundell1999,KaiserThorpe1997} and a maximum GRG linear size
cut off of 3 Mpc. An estimate of the predominant process of radiative
losses, obtained by separating the contributions of the inverse
Compton and synchrotron losses, shows that the ratio of the estimated
B$_{CMB}$/B$_{eq}$ increases with linear size, and IC losses dominate
the radiative losses in GRG \cite[]{Ishwara-Chandra1999}.

As argued by many authors, the observed physical characteristics
mentioned above could be the result of selection effects introduced by
the selection criteria or by biases due to the low sensitivity of
typical radio images.  The faintest regions of GRG are well detected,
even with a modest angular resolution, only with low frequency
interferometric observations. The low-frequency spectral index
information is crucial to derive the energy distribution of the
radiating electrons, and to study the energy transport from the
nucleus to the lobes in these exceptionally large radio sources.
Multifrequency spectral aging studies of GRG have been recently
presented by Jamrozy and collaborators
\citep{Jamrozy2004,Jamrozy2005,Jamrozy2008}. The median value for the
estimated spectral ages is 23-24 Myr. The injection spectral index
ranges from 0.55 to 0.88; it appears to increase with luminosity and
redshift but shows an inverse correlation with linear size.

In this paper we present a multifrequency spectral analysis of two
classical double giant radio galaxies 3C35 and 3C223. In Sect. 2 radio
observations and data analysis at 74 and 327\,MHz are described. In
Sect. 3 we present radio images of 3C35 and 3C223 at 74 and
327\,MHz. In Sect. 4 we show the spectral index maps and the spectral
analysis obtained by combining images at 74, 327, 608 and 1400
\,MHz. Results are discussed and summarized in Sect. 5.

\begin{figure*}[ht]
\begin{center}
\includegraphics[width=14cm]{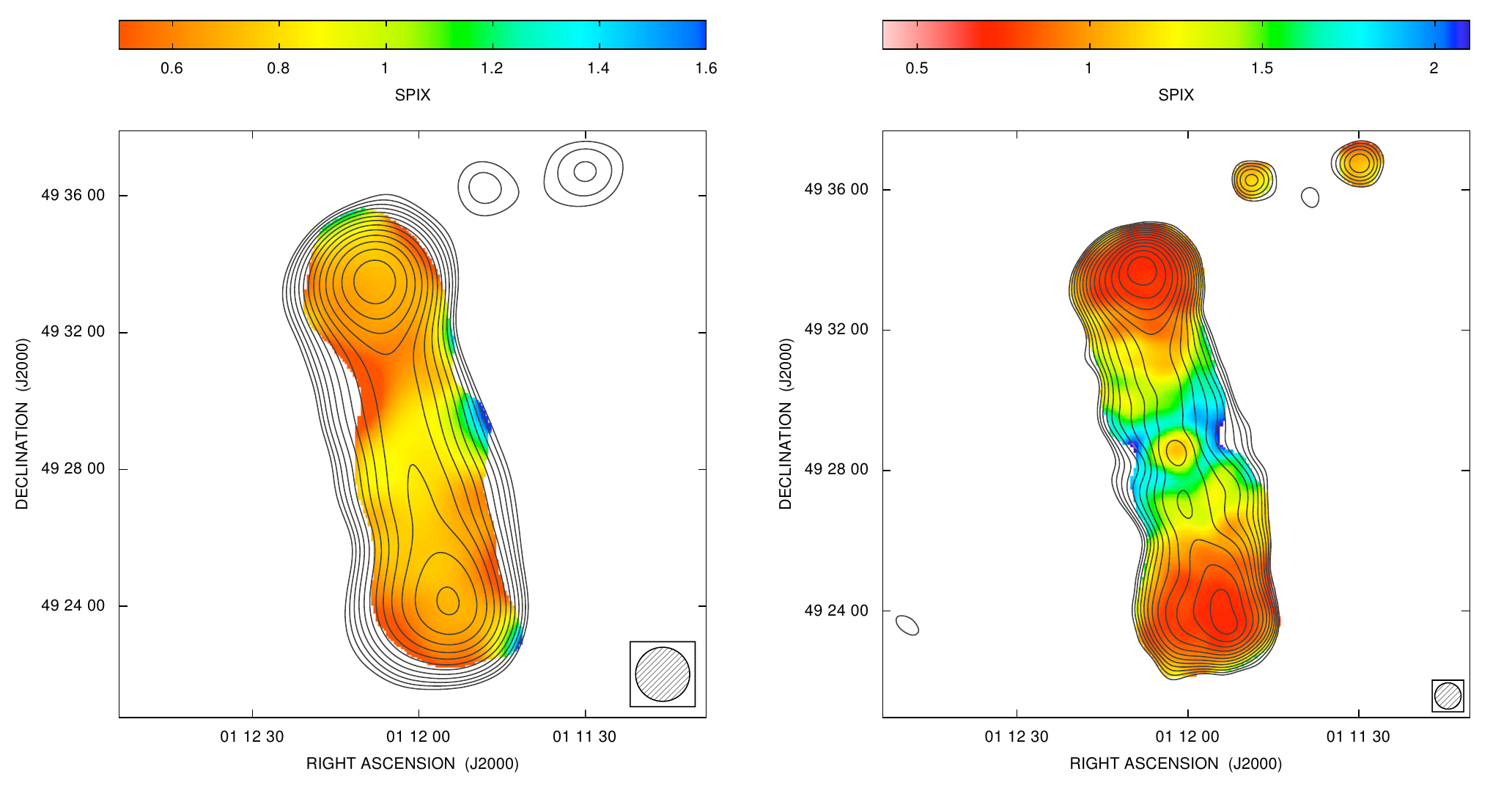}
\end{center}
\caption[]{ 3C35: Spectral index maps are shown in color; pixels
  whose brightness was below 3$\sigma$ have been  blanked. Contour
  levels are the radio brightness at 327\,MHz, start at (3$\sigma$) and are scaled by $\sqrt{2}$. {\it Left} Spectral index map between 74\,MHz and 327\,MHz, with a resolution of 95\arcsec$\times$95\arcsec. {\it Right} Spectral index map between 327\,MHz and 1.4\,GHz, with a resolution of 45\arcsec$\times$45\arcsec (the image at 1.4\,GHz was taken from the NVSS \cite[]{NVSS}).}
\label{3c35spixmap}
\end{figure*}

\begin{figure*}[ht]
\begin{center}
\includegraphics[width=14cm]{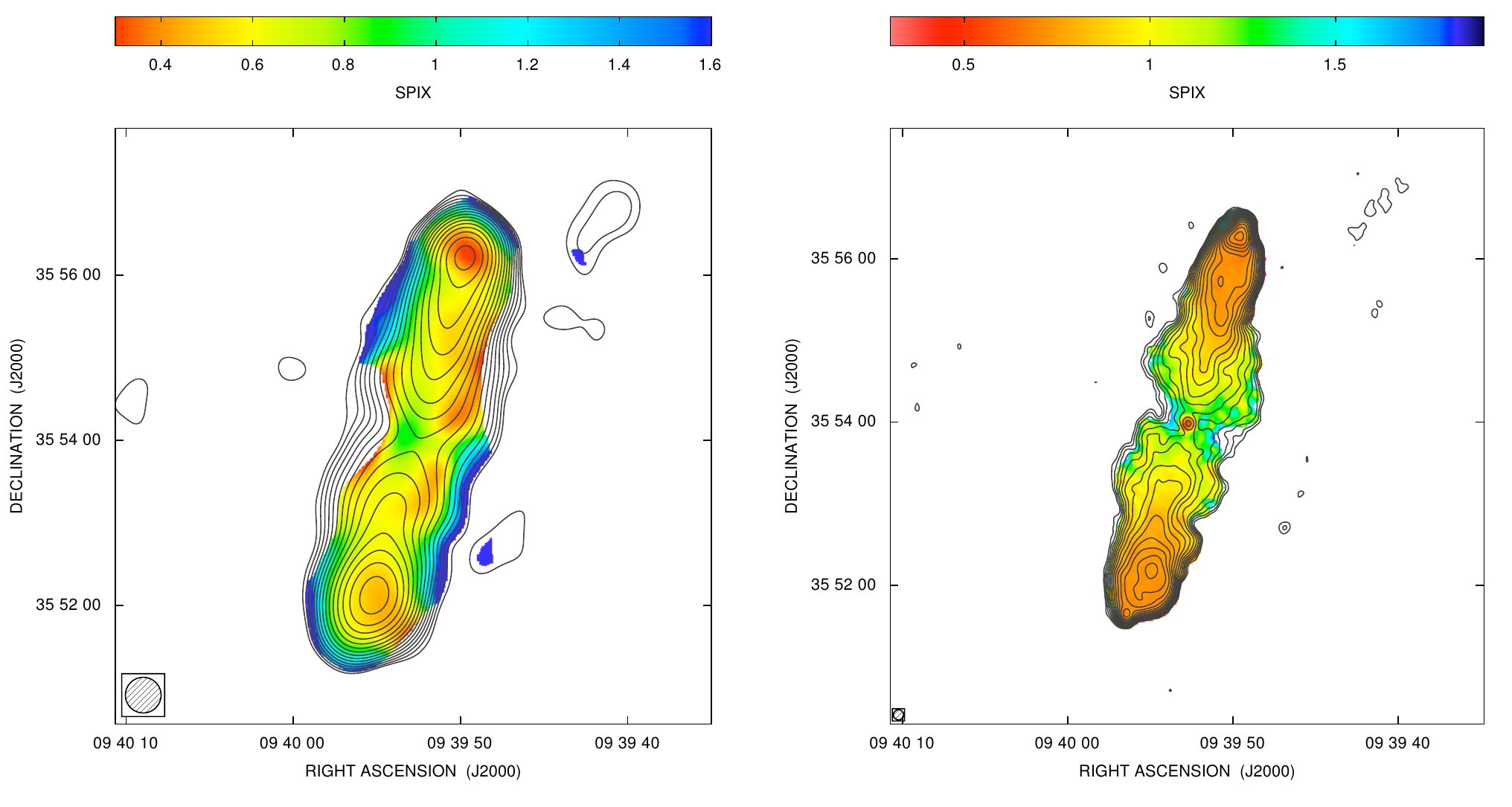}
\end{center}
\caption[]{3C223: Spectral index maps are shown in color; pixels whose brightness was below 3$\sigma$ have been
  blanked. Contour levels are the radio image at 327\,MHz , start at (3$\sigma$) and are scaled by $\sqrt{2}$. {\it Left} Spectral index map between 74\,MHz and 327\,MHz, with a resolution of 26\arcsec$\times$26\arcsec. {\it Right} Spectral index map between 327\,MHz and 1.48\,GHz, with a resolution of 7.5\arcsec$\times$7.5\arcsec.}
\label{3c223spixmap}
\end{figure*}

\section{Radio Data}
The two selected giant radio galaxies are 3C35 and 3C223. 3C35 is
included in the sample of 47 low redshift (z\,$<$\,0.4) GRG obtained
by \cite{Schoenmakers2001} using the Westerbork Northern Sky Survey
(WENSS) of the sky above $+30\degr$ of declination at 325\,MHz \cite[]{WENSS}.  The
criteria for the sample specified that a candidate GRG must have: {\it
i)} an angular size larger than 5 arcminutes, and {\it ii)} a distance
to the galactic plane of more than 12.5 degree.

\noindent
3C223 is included in a complete sample of large scale radio sources
selected by \cite{Leahy1991}.  The sources are drawn from a subset of
the complete radio sample with z less than 0.5 defined by
\cite{Laing1983}.

\noindent
3C35 and 3C223 have linear sizes of 950\,kpc and 780\,kpc
respectively, with the adopted cosmological parameters.

We observed these two GRG with the Very Large Array at 74 and 327 MHz
in several configurations, according to their angular dimensions, in
order to avoid the loss of flux. Observational parameters are
summarized in Table\,\ref{obs}.  Since observations were made at
slightly different frequencies, we will refer in the text simply to
observations at 74 and 327 MHz. The exact frequencies are reported in
section 3 and are used for estimates of the physical parameters.

\begin{figure*}[th]
\begin{center}
\includegraphics[width=14cm]{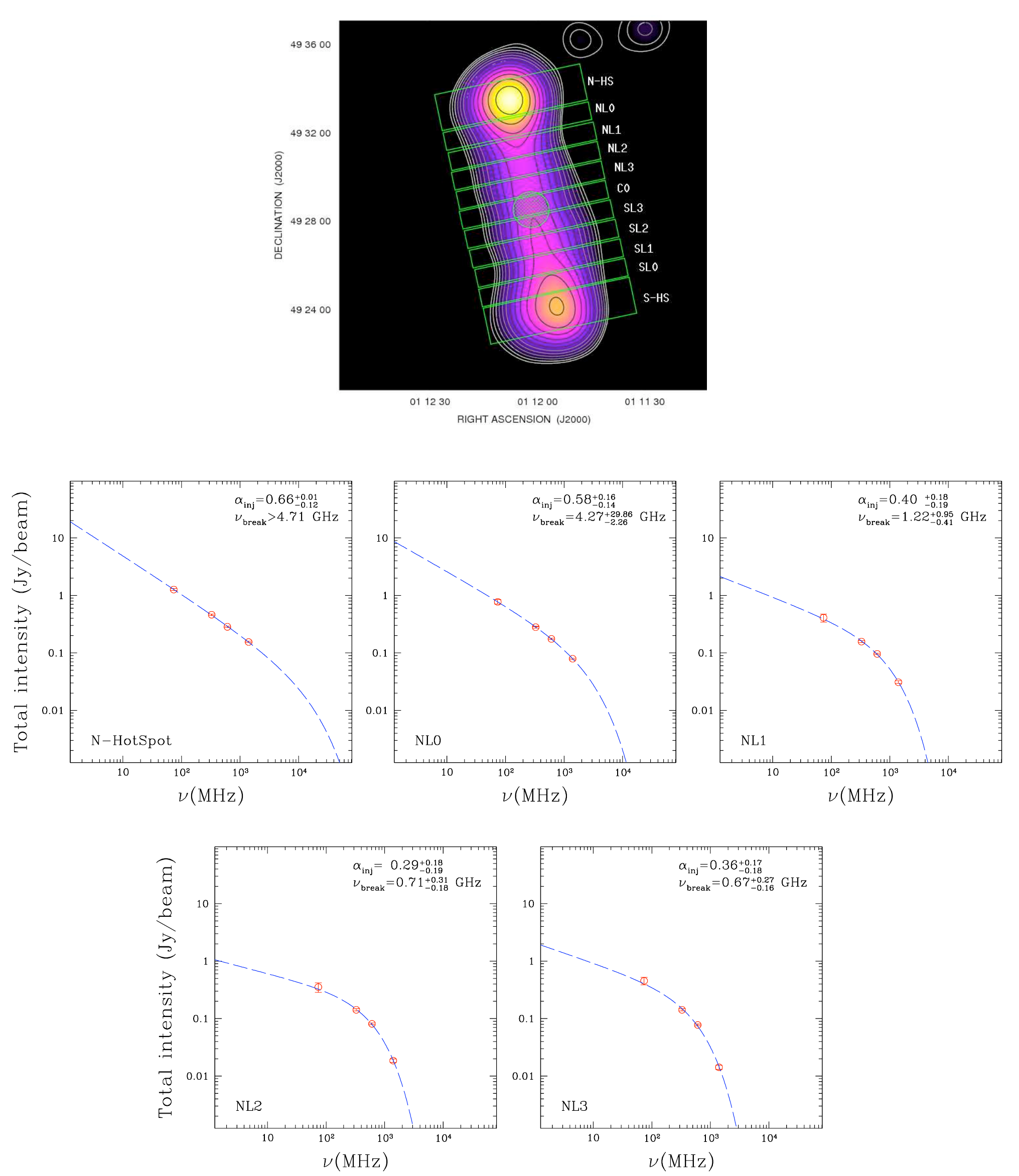}
\end{center}
\caption[]{{\bf a.} 3C35: Color map of the source (resolution
  95\arcsec) and the contours at 327 MHz; overlayed by the green array of
  boxes where the intensity measures have been
  taken. For the N lobe: plots represent the spectral shape of the source labeled
  according the positions. Points are measures taken at four
  frequencies, dashed lines are the fits with the synchrotron model.}
\label{3c35fitJP1}
\end{figure*} 
 
\setcounter{figure}{4}
\begin{figure*}[th]
\begin{center}
\includegraphics[width=14cm]{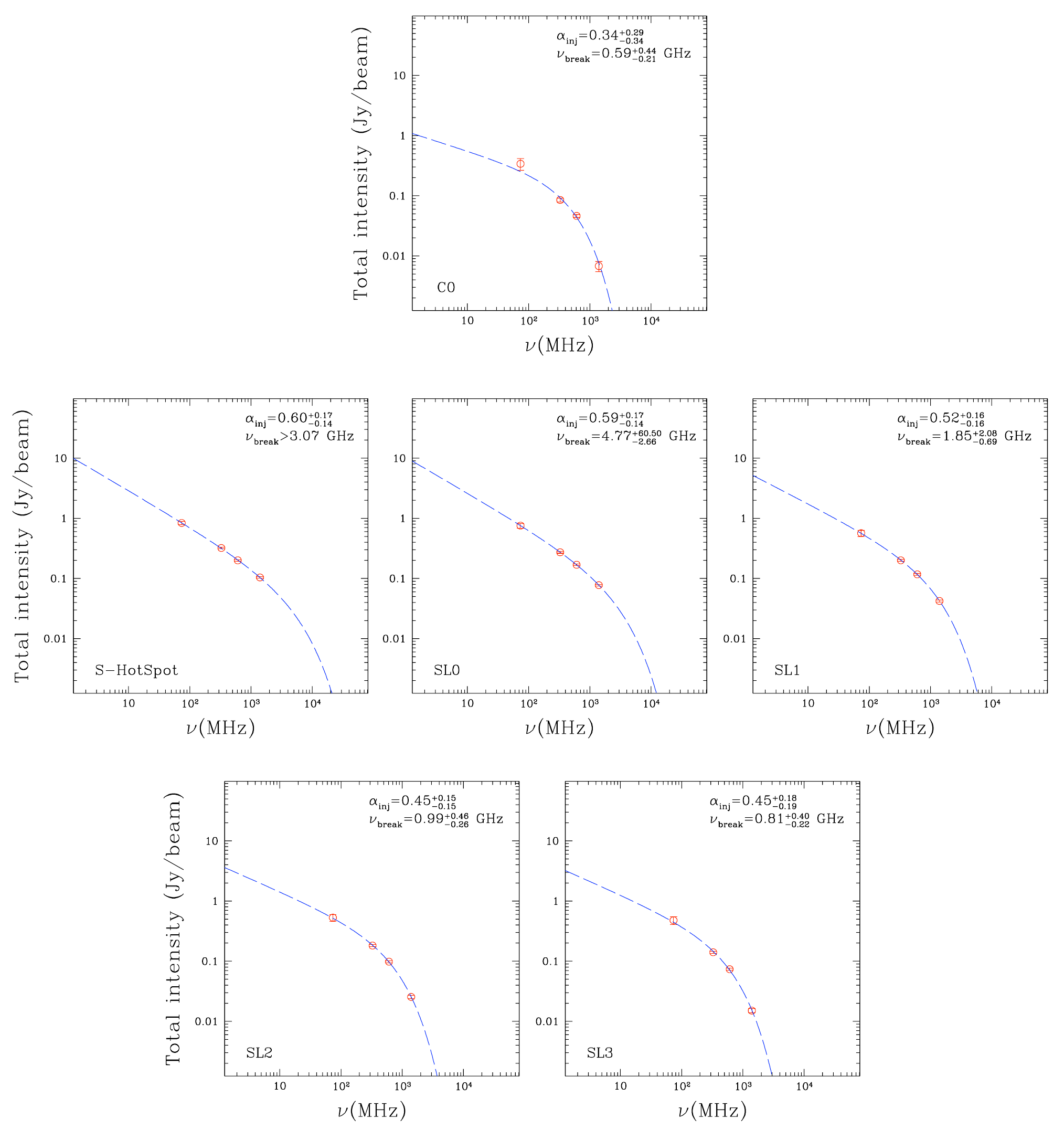}
\end{center}
\caption[]{{\bf b.}  Similar to Fig.\ref{3c35fitJP2} a for the S lobe
  of 3C35}
\label{3c35fitJP2}
\end{figure*} 

\begin{figure*}[th]
\begin{center}
\includegraphics[width=14cm]{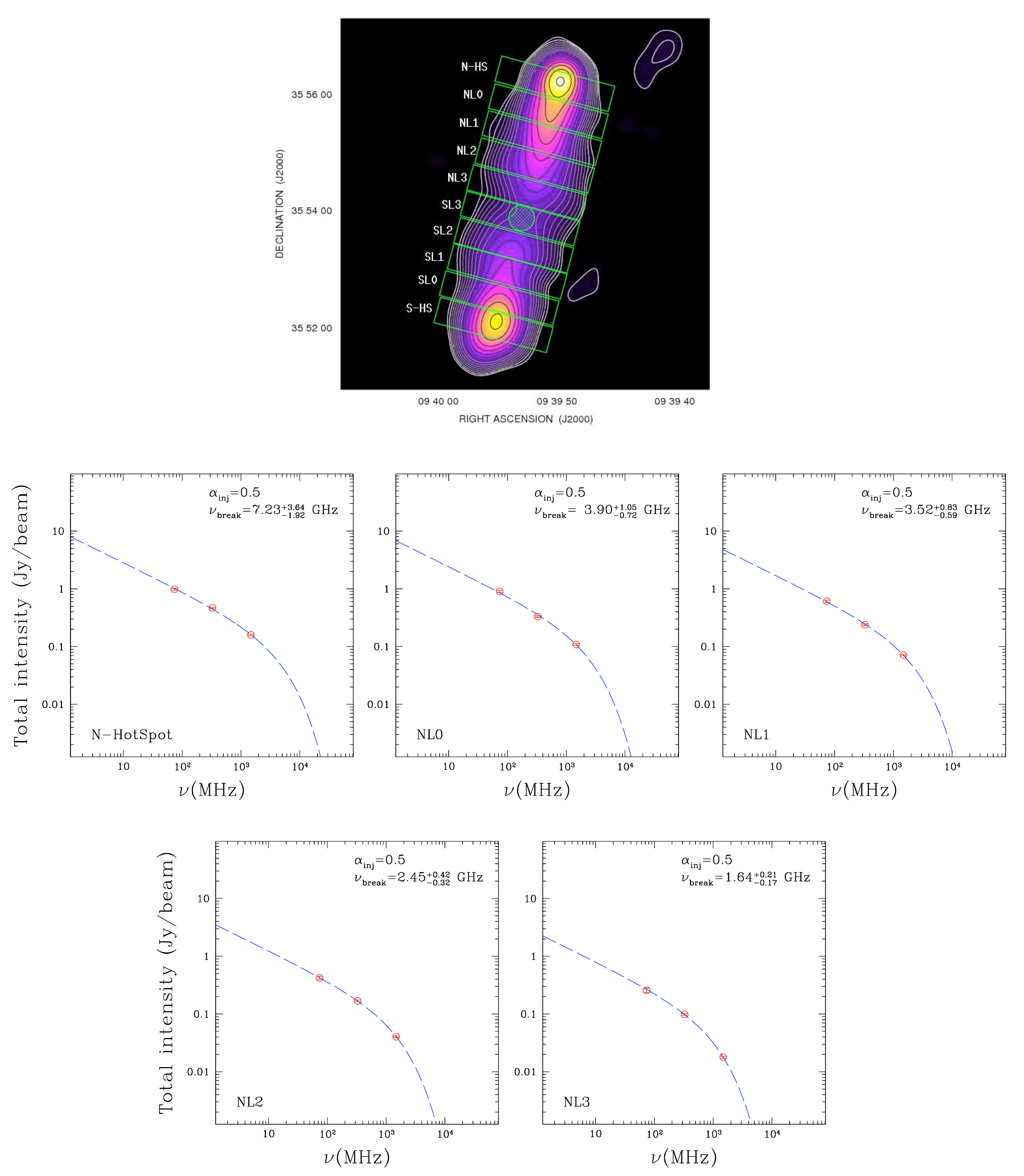}
\end{center}
\caption[]{{\bf a.} 3C223: Color map of the source (resolution
  26\arcsec) and the contours at 327 MHz ; overlayed the green array of
  boxes where the measures have been
  taken. For the N lobe: plots represent the spectral shape of the source labeled
  according the positions. Points are measures taken at three frequencies, dashed lines are the fits with the synchrotron model.}
\label{3c223fitJP1}
\end{figure*} 

\setcounter{figure}{5}
\begin{figure*}[th]
\begin{center}
\includegraphics[width=14cm]{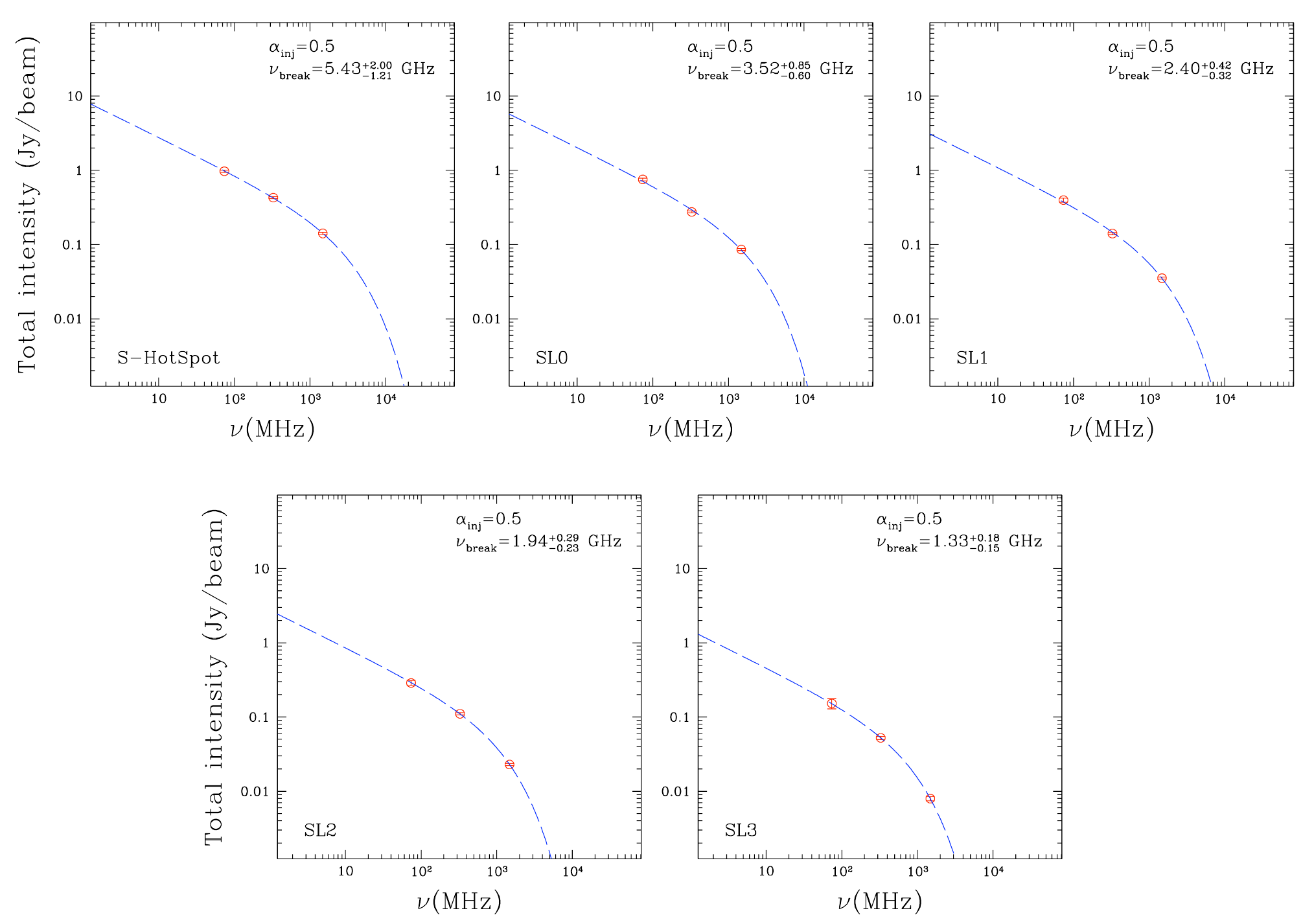}
\end{center}
\caption[]{{\bf b.} Similar to Fig.\ref{3c223fitJP1} for the the S lobe of 3C223}
\label{3c223fitJP2}
\end{figure*} 

\subsection{Observing strategy}

Radio Frequency Interference (RFI) strongly affects and corrupts the
data in low-frequencies observations. In order to permit the RFI
excision and to minimize the bandwidth smearing effect, the
observations were conducted in spectral line mode.  A difference has
been established in the nature of the RFI sources in the VLA system: at
74 MHz most of interference is caused by the 100 kHz oscillators in
the bases of each telescope, which generate harmonics at 100 kHz
intervals and produce the typical ``100 kHz {\it comb}''.  This kind
of RFI is ``easy'' to predict and eliminate.  In the 327 MHz band, the
internal electronics of the {\it VLA} gives rise to harmonics that are
multiples of 5 and 12.5 MHz; to avoid these narrow bandwidths
are used \cite[]{kassim1993}. On the whole, the values of rms
sensitivity attained at these frequencies are somewhat higher than the
expected thermal noise levels, because of the contribution of several
factors: confusion, broad-band RFI, and {\it VLA} generated RFI.

\subsection{Data Reduction}
Data were calibrated and reduced with Astronomical Image Processing
System (AIPS). Since the calibration procedures are different for 74
MHz and 327 MHz, the following sections describe the methods employed
separately.

\subsubsection{74 MHz}
The 74 MHz data were calibrated and imaged following the same
procedure used for the VLA Low-frequency Sky Survey (VLSS) as
described carefully in \cite{Namir2007} and \cite{Aaron2007}.

\noindent
Both for 3C35 and 3C223 we made the amplitude and bandpass calibration
using a model of Cygnus A\footnote{Available from
http://lwa.nrl.navy.mil/tutorial/VLAmodels}.

Careful data editing was used to excise the RFI \cite[]{Lane2005}; the
percentage of flagged data at the end of this process was about 13$\%$
for both sources.  Before the final imaging data were averaged to 8
frequency channels with a resolution of 170.9 kHz.

To produce the final images we used the so called "field-based
calibration" method developed for the VLSS \cite[]{Cotton2004}. The
field-based calibration procedure makes a correction for the low-order
in the ionospheric terms and performs a ''wide-field imaging'' (kindly
provided by W. D. Cotton). The offsets of the apparent positions of
the NRAO VLA Sky Survey (NVSS) sources \cite[]{NVSS} from their
expected positions were computed at time intervals of 2 minutes and
corrected in the visibility data. Some data with too large correction
were removed for 3C35.  We corrected the final images for the primary
beam effect.

The final image of 3C35 obtained with VLA data in B configuration has an rms 
sensitivity of $\sim$\,95\,mJy/beam (93\arcsec$\times$64\arcsec).   

For the source 3C223 we produced a high resolution image in A
configuration, with rms sensitivity $\sim$\,43\,mJy/beam
(25\arcsec$\times$24\arcsec).  A low resolution image was also made
with data from the VLA in B configuration; it has an rms sensitivity
$\sim$\,98\,mJy/beam (83\arcsec$\times$73\arcsec).  The image obtained
with the combination of A and B configurations data has a sensitivity
of 40\,mJy/beam (26\arcsec$\times$25\arcsec). The sensitivity is
different from the theoretical noise because of side-lobe confusion
from other sources in the beam.

\subsubsection{327 MHz}
We made the amplitude and bandpass calibration with the sources 3C48 and 3C286 
respectively for 3C35 and 3C223.

After careful data editing to remove RFI, 10$\%$ of data were flagged in 3C35
and 3$\%$ for 3C223.

Before imaging, the data of 3C35 were averaged to 5 channels with a
resolution of 488.3 KHz, and the 3C223 data were averaged to 6
channels with a resolution of 781.3 kHz.

The data were mapped using a wide-field imaging technique, which
corrects for distortions in the image caused by the non-coplanarity of
the VLA over a wide field of view (the "3-D effect" included in the
AIPS task IMAGR).  A set of small overlapping maps was used to cover
the central area of about $\sim\,1.5\degr$ in radius
\cite[]{Cornwell1992}.  However, at this frequency confusion lobes of
sources far from the center of the field are still present. Thus, we
also imaged strong sources in an area of about $\sim\,60\degr$ in
radius, based on positions in the NVSS catalog.  All these ``facets''
were included in the CLEAN and used for several loops of
self-calibration \cite[]{Perley1999}.  The data for each observation
and configuration were calibrated, imaged and then combined.  We
corrected the final images for the primary beam effect.

In particular, for 3C35, we obtain a high resolution image at 327\,MHz
with data from the VLA in B configuration; the rms sensitivity is
$\sim$\,1.3\,mJy/beam (23\arcsec$\times$17\arcsec). A low resolution
image was made from VLA observations in C configuration, with an rms
sensitivity of $\sim$\,2.3\,mJy/beam (55\arcsec$\times$50\arcsec) .
By combining the B configuration data and data from IF 1 of the C
configuration, we improved the uv-coverage and sensitivity; the rms in
the combined image is $\sim$\,1.0\,mJy/beam
(27\arcsec$\times$21\arcsec).

We obtained a high resolution image of 3C223 with VLA observations in
A configuration; the rms sensitivity is $\sim$\,0.7\,mJy/beam
(6\arcsec$\times$5\arcsec).  A low resolution image with a sensitivity
of $\sim$\,1.3\,mJy/beam was made from VLA data taken in B
configuration (19\arcsec$\times$16\arcsec).  To improve the
uv-coverage and the sensitivity we combined the datasets.  The final
rms sensitivity is $\sim$\,0.6\,mJy/beam (7\arcsec$\times$6\arcsec).
The final sensitivity differs from the theoretical noise due to
classical confusion.

\section{Results}

\subsection{3C35}

The source 3C35 is a classical double radio source with a regular FRII
structure \cite[]{Fanaroff1974}; its principal characteristics are
listed in Table \ref{source}. It has been previously studied at
frequencies of 608 MHz, 1.4 and 5 GHz with the Westerbork Synthesis
Radio Telescope (WSRT) (\citealt{vanBreugel1982};
\citealt{Jaegers1987}; \citealt{Schoenmakers2000}).

The VLA radio images at 74 and 327 MHz of the radio galaxy 3C35 are
shown respectively in the left and in the right panel of
Fig.\,\ref{3c35cntrnat}; sensitivities and resolutions are listed in
Table \ref{obs}.  On the left of Fig.\,\ref{3c35cntrnat} is the image
at 73.8 MHz obtained with VLA in B configuration. In the contours map
the regular double-lobe structure of the source is distinctly clear,
as well as the presence of the two hot-spots. The emission is stronger
at the head of the Northern lobe (N lobe).

On the right panel of Fig.\,\ref{3c35cntrnat} the 327 MHz image is
shown. The image is obtained by combining the data of B and C configurations at a
frequency of 327.4 MHz. The high and the low resolution images (not
shown in this paper) have been obtained with the B and the C
configurations of the VLA respectively. The image at 327\,MHz confirms that in the N
lobe the radio emission is stronger than in the Southern lobe (S
lobe).  The hot spot South (S hot spot) is slightly shifted with respect
to the axis of symmetry of the source. Based on the images at 5 GHz,
\cite{vanBreugel1982} claimed that the S hot spot could be a
double. Considering the image at 327 MHz, it seems
that the second weak ``hot spot'' is more likely to be a knot of the
jet. In the image the core and surrounding low
brightness emission have both been detected.

\subsection{3C223}
The host galaxy of 3C223 is in a group of 12 small galaxies
\cite[]{Baum1988}. It is a typical double radio source with a regular
FRII structure \cite[]{Fanaroff1974}. It has been previously studied
in with the WRST at 608 MHz, 1.4 and 5 GHz (\citealt{Hogbom1979};
\citealt{vanBreugel1982}; \citealt{Jaegers1987}) and at high
resolution with the VLA at 1.4 GHz \cite[]{Leahy1991}. The general
characteristics of this source are presented in Table\,\ref{source}.

The images at 74 and 327 MHz shown in figure\,\ref{3c223cntrnat} were
obtained by VLA observations using A and B configurations. The
sensitivities and resolutions are listed in Table \ref{obs}.

The 74 MHz image is shown on the left panel of
Figure\,\ref{3c223cntrnat}. A low resolution image has been obtained with the VLA in B
configuration  while the high resolution image has been made with the A
configuration (both images are not shown in this paper). To improve the uv-coverage and the
sensitivity, we combined the A and B configuration data. The combined
image, shown on the left of Fig.\,\ref{3c223cntrnat}, is made at an observed frequency of
73.8 MHz.  

\noindent
As can be observed in the higher brightness contours of the image, 
the morphology of the source preserves the FRII
structure. An extended low brightness structure is easily visible to
the West of the southern lobe and some faint extended emission seems
also to be present to the East of the northern lobe. This diffuse
structure shows a different orientation axis with respect to that of
the active lobes. The origin of this structure is not clear, it could
be the remnant of former radio emission, or a relic lobe. A hint of
this low brightness emission is present at the same position in the
existing image at 1.4\,GHz \cite{Leahy1991}.  The estimated spectral
index of this low brightnesses radio structure is $\alpha\,\sim\,$1.3
(obtained from the spectral index map between 1.4\,GHz and 74\,MHz not
shown here); this is conspicuously steeper than the average value
measured for the whole radio galaxy (Section 3.3 and
Tab.\ref{223flux}).

The image at 327 MHz is shown on the right panel of
Figure\,\ref{3c223cntrnat}.  An high resolution image at
327.3 MHz has been obtained with the VLA in A configuration, while the low
resolution image, obtained at the observing frequency of 328.9 MHz
with the B configuration (both images are not shown in this paper). In the right panel of
Figure\,\ref{3c223cntrnat} are presented the contours of the image obtained by combining the
A and B configuration data at the frequency of 327.3 MHz.

\noindent
Unlike the images at 608 MHz \cite[]{vanBreugel1982}, the core has
been clearly detected at 327 MHz thanks to the high resolution
achieved. High resolution obtained in the image at 327 MHz allows to
confirm the peculiar ``V'' shape of the N hot
spot, previously seen in high-resolution images at 1.4\,GHz \cite[]{Leahy1991}. Moreover, at
327 MHz, as well as in the high resolution images at 1.4 and 5 GHz
(\citealt{vanBreugel1982}; \citealt{Leahy1991}), the S hot spot seems
embedded in the lobe, at the end of which a protuberance is detected.

\subsection{Spectral index distribution}
By combining the new images at 74\,MHz and 327\,MHz with those at
1.4\,GHz available in the literature, we obtained the spectral
index\footnote{S($\nu$)\,$\propto$\,$\nu^{-\alpha}$}
distributions of the two radio galaxies 3C35 and 3C223. Figures
\ref{3c35spixmap} and \ref{3c223spixmap} show the spectral index maps
of 3C35 and 3C223, respectively. Both figures show on the left the
spectral index maps between 74\,MHz and 327\,MHz, while on the right
are those between 327\,MHz and 1.4\,GHz.  In the range between 74\,MHz
and 327\,MHz (95\arcsec$\times$95\arcsec), the spectral index values
of 3C35 vary from $\alpha\,\sim$\,0.65\,$\pm$\,0.04, in the main parts
of the source, up to $\alpha\,\sim$\,0.84\,$\pm$\,0.06 in the region
near the core.

\begin{table}
\caption{Resolution and rms of the images used for the spectral index
  maps.}
\label{35fitpar}
\begin{center}
\begin{tabular}{cccc}
\hline           
name&$\nu$&resolution&rms\\
&MHz&\arcsec$\times$\arcsec&mJy/beam\\
\hline
\noalign{\smallskip}
3C35&74&95$\times$95&92.4\\
&327&95$\times$95&7.0\\
&327&45$\times$45&2.1\\
&1400&45$\times$45&0.5\\
\noalign{\smallskip}
\hline
\noalign{\smallskip}
3C223&74&26$\times$26&43.6\\
&327&26$\times$26&3.5\\
&327&7.5$\times$7.5&0.7\\
&1480&7.5$\times$7.5&0.1\\
\noalign{\smallskip}
\hline
\end{tabular}
\end{center}
\end{table}

The spectral index between 327\,MHz and 1.4\,GHz
(45\arcsec$\times$45\arcsec) for 3C35 varies more than the lower
frequency index. In the region of the head of the lobes, $\alpha$ is
about 0.72\,$\pm$\,0.01, while it reaches values of
$\sim$\,1.6\,$\pm$\,0.04 in the inner region of the lobes around the
core. The image at 1.4\,GHz used to obtain the spectral index map has been
taken from the NVSS \cite[]{NVSS}. The morphology of this source at
1.4\,GHz recalls the one observed at 327\,MHz.

For the radio galaxy 3C223, the spectral index distribution between 74
and 327\,MHz (26\arcsec$\times$26\arcsec) is slightly more patchy than
that of 3C35. The average value of $\alpha$ is about
0.60\,$\pm$\,0.03. In the North hot spot the spectral index reaches
values of $\sim$ 0.40\,$\pm$\,0.01, while in the low brightness
regions of the lobes the spectral index steepens up to
$\alpha\,\sim$\,1.67\,$\pm$\,0.02.

\noindent
The spectral index map between 327\,MHz and 1.48\,GHz
(7.5\arcsec$\times$7.5\arcsec) for 3C223 shows that the spectral index
increases from $\alpha\,\sim$\,0.72\,$\pm$\,0.01 in the region of the
head of the lobes, up to values of $\alpha\,\sim$\,1.32\,$\pm$\,0.09
in the inner regions of the lobes near to the core.  The image used at
1.48\,GHz was made from VLA archive data \cite[]{Leahy1991}.

\subsection{Measure and estimate of physical parameters}
For both sources the flux density at 74 and 327\,MHz was measured in
the same regions.  The values listed in Table\,\ref{35flux} and
Table\,\ref{223flux} indicate the flux of the entire source, of the
two lobes separately and of the core when detected. For these regions
we also calculated the spectral index between 74 and 327\,MHz and the
equipartition magnetic field (Tables\,\ref{35flux} and
\ref{223flux}). The measured total flux densities at 327\,MHz have been compared with
  the measurements of the WENSS, these are in agreement within the errors for both sources.

\begin{table}
\caption{3C35 Flux densities, spectral indices and equipartition magnetic fields.}
\label{35flux}
\begin{center}
\begin{tabular}{ccccc}
\noalign{\smallskip}
\hline
\noalign{\smallskip}

& Total& N lobe & S lobe & Core  \\
\hline
\noalign{\smallskip}
F$_{73.8 \rm MHz}$(Jy)& 20.9$\pm$0.7 &10.8$\pm$0.4 & 10.2$\pm$0.5 \\
\hline
\noalign{\smallskip}
F$_{327.4 \rm MHz}$(Jy)&7.5 $\pm$0.2 &3.9$\pm$ 0.1 & 3.6$\pm$0.1&0.171$\pm$0.004\\
\hline
\noalign{\smallskip}
$\alpha_{74}^{327}$&0.7&0.7&0.7& \\
\noalign{\smallskip}
\hline
\noalign{\smallskip}
B$_{eq-\nu}$($\mu$G)&0.72&0.73&0.71& k=1\\
&0.59&0.59&0.58& k=0\\
\hline
\noalign{\smallskip}
B$_{eq-\gamma}$($\mu$G)&1.03&1.04&1.02& k=1\\
&0.85&0.86&0.84& k=0\\
\hline
\hline
\noalign{\smallskip}
\end{tabular}

\end{center}
\end{table}

\begin{table}
\caption{3C223 Flux densities, spectral indices and equipartition magnetic fields.}
\label{223flux}
\begin{center}

\begin{tabular}{ccccc}
\noalign{\smallskip}
\hline
\noalign{\smallskip}
& Total& N lobe & S lobe & Core  \\
\hline
\noalign{\smallskip}
F$_{\rm 73.8 MHz}$(Jy)&30.2$\pm$1.0&16.2$\pm$0.6 &14.0$\pm$0.5 \\
\hline
\noalign{\smallskip}
F$_{\rm 327.3 MHz}$(Jy)&11.7$\pm$0.4&6.4$\pm$0.2 &5.2$\pm$0.2&0.031$\pm$0.002\\
\hline
\noalign{\smallskip}
$\alpha_{74}^{327}$& 0.6&0.6&0.7& \\
\noalign{\smallskip}
\hline
\noalign{\smallskip}
B$_{eq-\nu}$($\mu$G)&1.28&1.31&1.26& k=1\\
&1.05&1.07&1.03& k=0\\
\hline
\noalign{\smallskip}
B$_{eq-\gamma}$($\mu$G)&1.58&1.62&1.75& k=1\\
&1.31&1.34&1.45& k=0\\
\hline
\hline
\noalign{\smallskip}
\end{tabular}
\end{center}
\end{table}

\noindent
The zero-order estimate of the magnetic field strength, averaged over
the entire source volume, can be derived under the assumption of
classical equipartition.  Here, the radio source is in a minimum
energy condition and the relativistic particle energy density equals
the magnetic field energy density.  In the framework of the
equipartition hypothesis, the magnetic field can be determined from
the radio synchrotron luminosity and the source volume.  We estimated
the equipartition magnetic field assuming a magnetic field entirely
filling the radio source in a range of frequencies in which the
synchrotron luminosity is calculated from a low frequency cutoff of 10
MHz to a high frequency cutoff of 10 GHz. The volume averaged magnetic
fields were evaluated within a cylinder.  Then we considered two
cases: a) one in which the energy is equally divided between relativistic
protons and electrons--positrons, and the ratio between protons
and electrons k=p/e will be k=1, and b) one case in which all the energy
is provided by a plasma of relativistic electrons--positrons and k=0.

\noindent
In our case we consider a power-law injection spectrum with index
$\delta$, therefore, since  $\alpha = (\delta - 1)/2$, we adopted the
measured spectral index $\alpha$ for the estimate of the magnetic
field.

\noindent
Assuming a low-frequency cut-off of 10\,MHz in the luminosity
calculation is equivalent to assuming a low-energy cut-off of
$\gamma_{min}\sim 2000$ in the particle energy spectrum (B$_{eq-\nu}$
in Tables \ref{35flux} and \ref{223flux}). Therefore, in our estimates
of $B_{eq}$ we adopted the revised formalism (\citealt{Brunetti1997},
\citealt{Beck2005}) assuming a low-energy cut-off of
$\gamma_{min}=100$ in the particle energy distribution rather than a
low-frequency cut-off in the emitted synchrotron spectrum
(B$_{eq-\gamma}$ in Tables \ref{35flux} and \ref{223flux}).

\noindent
For 3C35 we considered a cylinder of 190 kpc of radius and height
$\sim$ 1 Mpc.  The adopted spectral index of the electron energy
spectrum, between 74 and 327 MHz (Table\,\ref{35flux}) is
$\alpha\,\simeq\,0.7$, which yields $\delta\,=\,2.4$. The volume of
the two lobes respectively is one half of the total volume. The
measure of the fluxes includes the emission of the hot-spots. The
estimated equipartition magnetic field strength values are listed in
Table\,\ref{35flux}.

\noindent
For the radio source 3C223, we assumed a cylinder with radius of
$\sim$\,200\,kpc and height $\sim$\,900\,kpc.  The spectral index of
the electron energy spectrum of the entire source and of the North
lobe is $\delta\,=\,2.2$ which corresponds to
$\alpha^{327}_{74}$\,$\simeq$\,0.6, while for the South lobe we
used $\delta$\,=\,2.4 which corresponds to
$\alpha^{327}_{74}$\,$\simeq$\,0.7. The contribution of the
hot-spot emission was included in these measures. The resulting
equipartition magnetic fields strength are presented in Table\,\ref{223flux}.

One of the most important physical properties of GRG, which
  distinguish them from compact and powerful sources like e.g. CygA,
  is that the equipartition magnetic field is far below the inverse
  Compton equivalent field over most of the lobes (see Tab. 4 \&
  5). This means that the electron energy losses are largely
  dominated by the inverse Compton scattering of the CMB photons,
  which can be assumed fairly uniform and isotropic. 
Thus, the location of the break energy is mostly unaffected by
eventual gradients of the magnetic field in the lobe. 
However, very strong negative magnetic field gradients from the
hot-spot to the core, if any, could affect the location of the break
frequency and mimic the effect of aging, i.e. the source would appear
older than what really is. Although, at least in the case of 3C223, we
know from the X-ray data that the equipartition magnetic field is
correct to within a factor of 2.
However, in the hot-spots the situation is different and the magnetic
field could be significantly higher than $B_{IC}$, where $B_{IC}$
  is the magnetic field directly estimated by using X-ray and radio data. 
On the other hand, in this regions radiative losses are balanced by re-acceleration and
injection of new particles to form what we define as "the zero-age
injection spectrum" \cite[see the discussion in][]{Carilli1991}. 
Unfortunately, we do not have enough resolution in our images to resolve
the hot-spots in our GRGs and thus to address this issue further in details.

\begin{figure}[]
\begin{center}
\includegraphics[width=9cm]{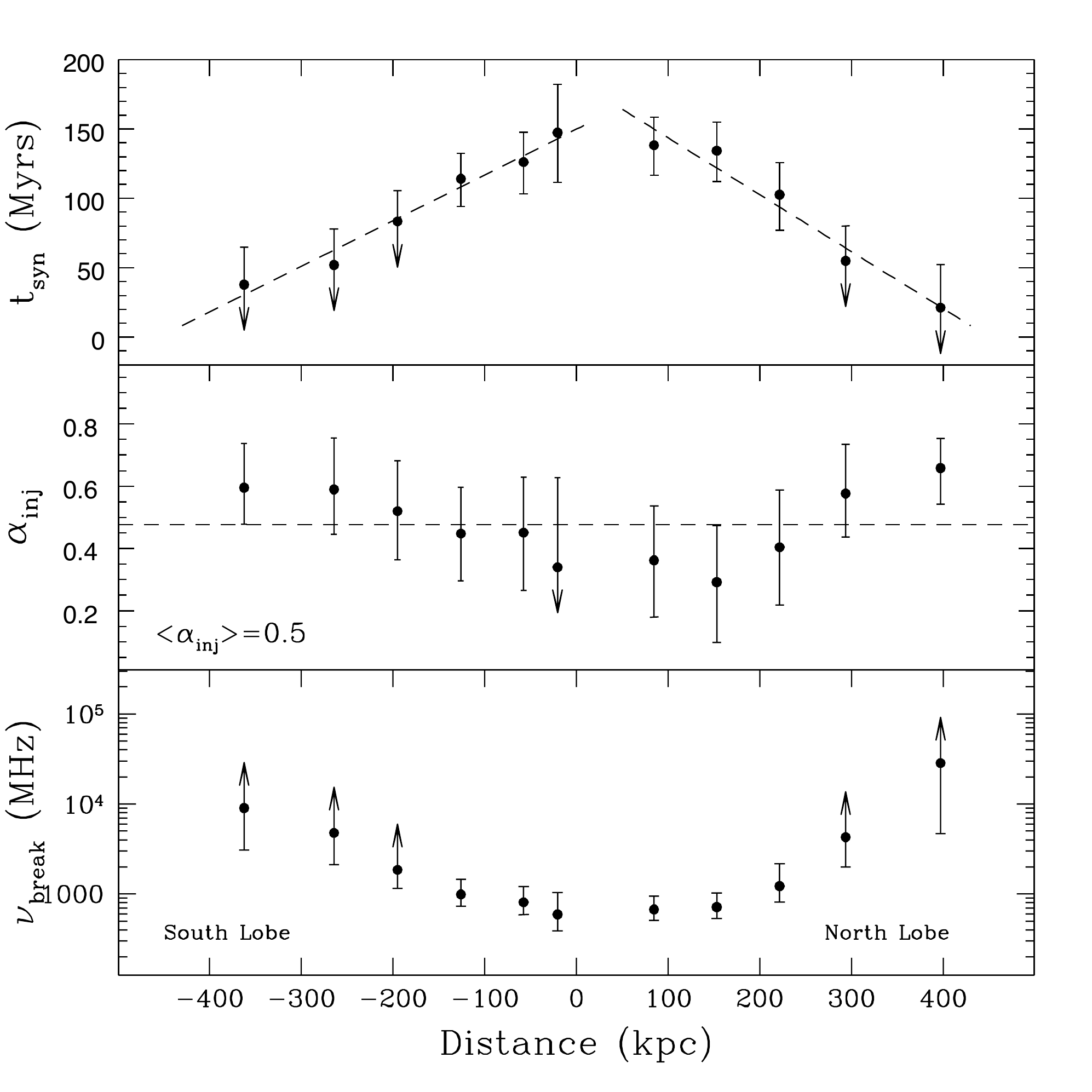}
\end{center}
\caption[]{3C35: plotted with 
respect to the distance from the core are: the fitted values of $\nu_{break}$ ({\it bottom}); the 
fitted values for $\alpha_{inj}$ ({\it center}); the estimated synchrotron ages ({\it top}).}
\label{3c35fitres}
\end{figure} 

\begin{figure}[]
\begin{center}
\includegraphics[width=9cm]{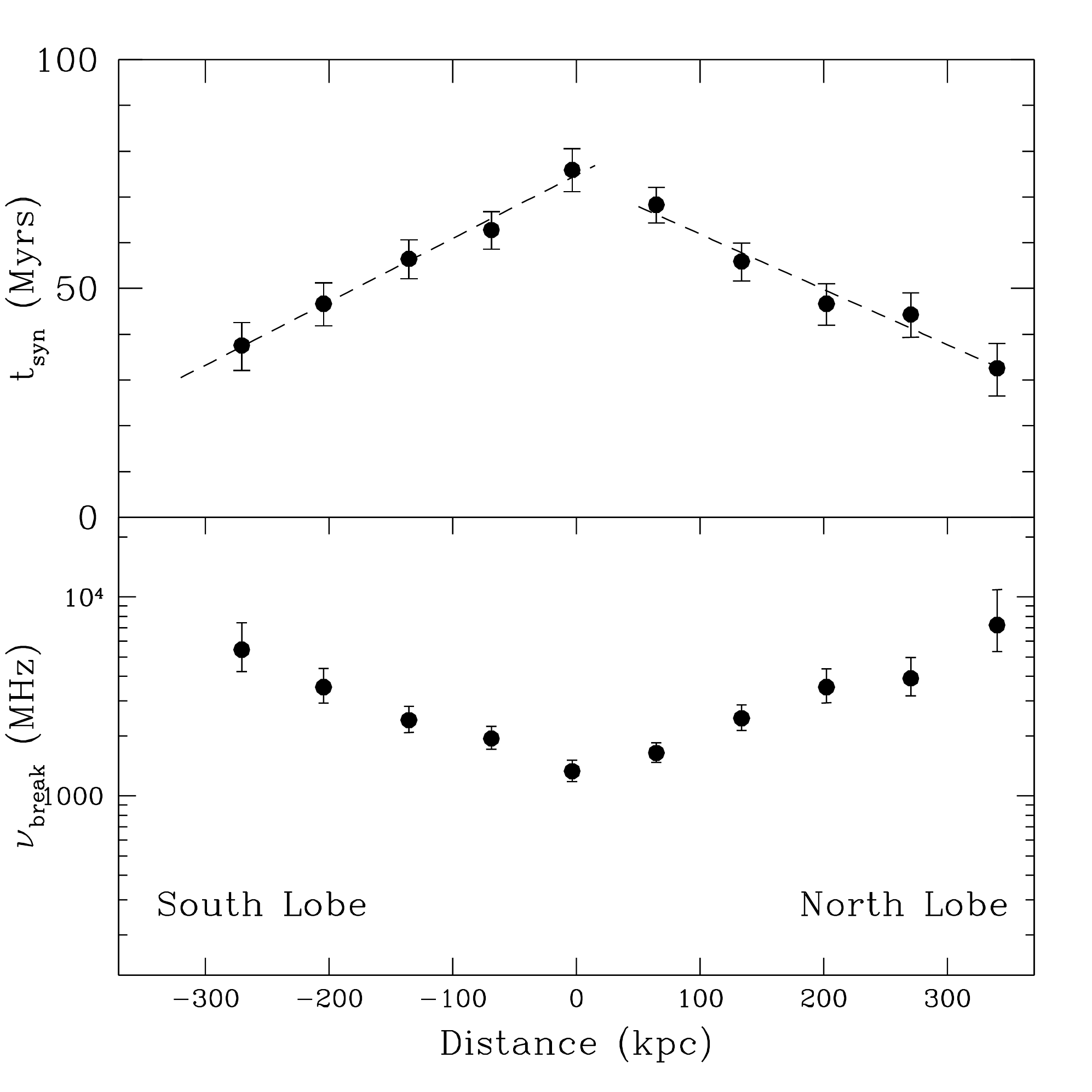}
\end{center}
\caption[]{3C223: plotted with 
respect to the distance from the core: the fitted values of $\nu_{break}$ ({\it bottom}) and 
the estimated synchrotron ages ({\it top}).}
\label{3c223fitres}
\end{figure} 

\section{Synchrotron Model}
\subsection{Fitting the spectral shape}
\label{multifreq}
The extended size of the two sources and the resolution reached in our
images allowed us to describe at different frequencies the variation
of the total intensity, from the hot-spot, where the particles are
injected, up to the inner regions of lobes, where the older less
energetic electrons are supposed to be located.  We assumed that the
particles are injected by the jets in the intergalactic medium, in a
certain epoch $t_0$, with a power-law energy spectrum:
\begin{equation}
N(\eps)\propto N_0 \eps^{-\delta_{inj}};
\label{N(e)}
\end{equation}
then the jets move further injecting particles in another region. This
continues up to the hot spots where particles are injected at the
current epoch.  The relativistic particles emit a synchrotron
radiation and the spectrum is still a power-law with a spectral index
$\alpha_{inj} = (\delta_{inj} - 1)/2$.  Particles are affected by
radiative losses via synchrotron and inverse Compton processes,
therefore the radio spectrum moves away from the original power-law
with a break at high frequencies.

The JP model \cite[]{JP} was used in this analysis. In this model the
pitch angle scattering is very efficient; indeed the isotropization
occurs on a time scale much shorter than the radiative
time-scale. Because the pitch angle is continuously randomized, each
particle of the electron population can assume all possible
orientations with respect to the magnetic field.  In the framework of
the JP model, the break frequency $\nu_{break}$ is a time-dependent
function, that can be estimated by fitting the curvature of the radio
spectra. If there is no expansion and the magnetic field is constant,
the break frequency depends on the elapsed time since the injection of
the particles according to this formula \cite[]{Slee2001}:
\begin{equation}
\label{tsyn}
t_{syn}=1590{\frac{B_{eq}^{1/2}}{(B_{eq}^{2}+B^{2}_{\rm{CMB}})}}{\frac{1}{[ \nu _{break}(1+z)]^{1/2}}}~~~\rm Myrs
\end{equation}
where $t_{syn}$ is the radiative age of the source, $\nu_{break}$ is
measured in GHz, B$_{eq}$ and B$_{CMB}=3.25\,(1+z)^2$ are the
equipartition and inverse Compton equivalent magnetic fields
respectively, both measured in $\mu$G.

For each source we measured the total intensity for the N lobe and S
lobe in a grid of boxes placed along the source (top panels of the
Figures \ref{3c35fitJP1}a and \ref{3c223fitJP1}a).  For the source 3C35
we choose the size of the boxes of half of the beam size for the
  lobes in order to have an even number of nearly independent
  measures, this fine sampling does not affect the age estimate. 
Since the poor resolution, to avoid the breaking off of the hot-spots
in two boxes, the boxes of the two hot-spots have been selected of
about one beam size. For the source 3C223 boxes are about one beam size.

 The plots in the figures
\ref{3c35fitJP1}a and \ref{3c223fitJP1}a refer to the N lobes while
the \ref{3c35fitJP2}b and \ref{3c223fitJP2}b show the S lobes
respectively of the sources 3C35 and 3C223.  The contribution of the
emission from the core has been masked for both sources. Each plot
corresponds to a box as shown in the labels on the bottom-left
corner. The red dots represent the data, while the blue dashed lines
are the fit with the model.

The model used to fit the data is the JP model available in the
software package Synage++ \cite[]{MurgiaPhD}. For these two radio
sources, the choice of the JP model is justified by the fact that
since the magnetic field is low, the inverse Compton losses are as
important as synchrotron losses, therefore the CMB isotropises the
electron population.

The spectra of 3C35 have been obtained using cube images with
frequencies of 74, 327, 608 and 1400 MHz at the resolution of
95\arcsec. To perform this analysis we used the
608\,MHz WSRT image with a resolution of 
40\arcsec$\times$20\arcsec, PA=0\degr and rms 1.3 mJy/beam; the
1.4\,GHz image has been taken from
the NVSS. The free parameters in the fit are: the break frequency
$\nu_{break}$, the injection spectral index
$\alpha_{inj}$ and  the flux normalization, which is proportional to the integral along the line
of sight of the product $N_0 \times B^{1+\alpha}$. The resulting fitted parameters obtained for each box
and the reduced $\chi^2$ ($\chi^2$/ndf, where ndf is the
number of degrees of freedom) are listed in Tab.\ref{35fitpar}. 
The shape of the hot spots are well
described by power-laws; the estimated break frequencies are
$>$4.71\,GHz for the North 
hot spot and $>$ 3.07\,GHz for the South. The
injection spectral index is 0.66$_{-0.12}^{+0.09}$ and
0.59$_{-0.12}^{+0.14}$ for the North and for the South hot spots
respectively. In the inner part of the lobes $\nu_{break}$ is $\simeq$
700-800\,MHz. The fitted spectra are plotted in Fig.\ref{3c35fitJP1}a
and \ref{3c35fitJP2}b and summarized in Fig.\ref{3c35fitres} as a
function of the distance from the core. The central panel of
Fig.\ref{3c35fitres} shows a variation of the injection spectral index
along the source, the values range from 0.29 to 0.66; on average
$<\alpha_{inj}>\,\simeq$\,0.5.  A variable spectral index could
  be explained in terms of deceleration of the relativistic plasma
  along the jets, following the subsequently happening of multiple
  shocks \cite[]{Meli2008}. This could be observed with
  a steepening of the electron spectra and in parallel of $\alpha_{inj}$ with the aging of the
  source. Observations of the electron spectra from the terminal 
hot-spots to the lobes of the powerful FR-II radio galaxies showed that
these have not a single and universal power-law form \citep[]{Rudnick1994,Machalski2007}.  

Nevertheless, is should be considered that our error bars on $\alpha_{inj}$ are large,
and hence the injection spectral index could still be considered fairly 
constant around a value of $\simeq$\,0.5. Future observations at possibly higher sensitivity 
 are needed to confirm the $\alpha_{inj}$ trend along the lobes definitely.

The top panel of Fig.\ref{3c35fitres} shows the synchrotron age
of 3C35 calculated by using the formula \ref{tsyn}. The values used for
$\nu_{break}$ and B$_{eq\nu}$ (k=1) are listed respectively in
Tab. \ref{35fitpar} and \ref{35flux}. The synchrotron age calculated
for the source 3C35 ${138}_{-22}^{+20}$ Myr for the North lobe
  and ${147}_{-36}^{+35}$ Myr for the South lobe..  
Therefore, in agreement with \cite{Parma1999}, 3C35 can be considered an old source.  Moreover,
since from the plot we can see that the $t_{syn}$ increases linearly
with the distance from the hot-spots, we can conclude that the source
reached this size expanding with a constant velocity in the
extragalactic medium $v_{exp}\sim$\,0.011c.

In the analysis of 3C223 we used images at 74, 327 and 1400 MHz with a
resolution of 26\arcsec. The 1.4\,GHz image is from  \cite{Leahy1991}.
We used a fixed $\alpha_{inj}$\,=\,0.5, assumed according to the literature,
because the observational data were only for three frequencies, which
restricts to two the number of free parameters. The free parameters in
this case are the break frequency $\nu_{break}$ and the flux
normalization. The resulting fitted parameters obtained for each box
and the reduced $\chi^2$ are listed in Tab. \ref{223fitpar}.  In this case the spectral shape
and the 
of the hot spots drifts away from a power-law, a $\nu_{break}$ of
about 6.0-7.0\,GHz has been estimated. In the inner regions of the
lobes the spectrum steepens and the $\nu_{break}$ is
$\simeq$1.5\,GHz. The fitted spectra are plotted in
Fig.\ref{3c223fitJP1}a and \ref{3c223fitJP2}b and summarized as
function of the distance from the core in Fig.\ref{3c223fitres}.  

In the top panel of Fig.\ref{3c223fitres} we plot the synchrotron age
of 3C223 estimated from the formula \ref{tsyn}, and using the values
of $\nu_{break}$ and B$_{eq}$ in Tab. \ref{223fitpar} and
\ref{223flux}.  For the source 3C223 the $t_{syn}$  is ${68}_{-4}^{+4}$ for the North 
lobe and ${76}_{-5}^{+5}$ Myr for the South lobe. As
for 3C35, the age decreases linearly with the distance from the core,
so we can assume that the lobes advanced with a constant velocity $v_{exp}\sim$\,0.017c.

 The 5\,GHz archival data were not included in the spectral aging analysis since the 
 low-surface brightness emission near the core 
is not properly imaged by the interferometric observations at 5\,GHz due to the lack of
short-space baselines. 
However, it should be noted that the fit of the JP model is able to recover also a break frequency
significantly above the maximum frequency in our data set since the spectral shape departs from a pure power law
well before $\nu_{break}$. In fact, although in the JP model all particles have the same break energy, 
 the break frequency of the particles at small pitch angle, $\theta$, is also smaller:  $\nu_{break}(\theta)\propto sin\theta$.

\begin{table}
\caption{3C35: break frequency, alpha
    injection and reduced  $\chi^2$.}
\label{35fitpar}
\begin{center}
\begin{tabular}{cccc}
\hline           
Region & $\nu_{break}$& $\alpha_{inj}$ & $\chi^2_{red.}$ \\
& GHz& & \\ 
\hline
N-HotSpot&$>4.71$&${0.66}_{-0.12}^{+0.09}$ &0.3 \\
NL0&${4.27}_{-2.26}^{+29.86}$&${0.58}_{-0.14}^{+0.16}$ &0.01 \\
NL1&${1.23}_{-0.41}^{+0.41}$&${0.40}_{-0.19}^{+0.18}$& 0.5\\
NL2&${0.72}_{-0.18}^{+0.31}$&${0.29}_{-0.19}^{+0.18}$&0.6 \\
NL3&${0.67}_{-0.16}^{+0.27}$&${0.36}_{-0.18}^{+0.17}$& 2.8\\
C0&${0.59}_{-0.21}^{+0.44}$&${0.34}_{-0.34}^{+0.29}$ &4.2 \\
SL3&${0.81}_{-0.22}^{+0.40}$&${0.45}_{-0.19}^{+0.18}$ & 1.4\\
SL2&${0.99}_{-0.26}^{+0.46}$&${0.45}_{-0.15}^{+0.15}$ &0.2 \\
SL1&${1.85}_{-0.69}^{+2.08}$&${0.52}_{-0.16}^{+0.16}$& 0.05\\
SL0&${4.77}_{-2.66}^{+60.00 }$&${0.59}_{-0.14}^{+0.17}$& 0.01\\
S-HotSpot&$>3.07$&${0.59}_{-0.12}^{+0.14}$& 0.2\\
\hline
\noalign{\smallskip}
\end{tabular}
\end{center}
\end{table}

\begin{table}
\caption{3C223: break frequency and reduced  $\chi^2$.}
\label{223fitpar}
\begin{center}
\begin{tabular}{ccc}
\hline           
Region & $\nu_{break}$& $\chi^2_{red.}$ \\
& GHz& \\ 
\hline
N-HotSpot& ${7.23}_{-1.92}^{+3.64}$&1.3\\
NL0& ${3.90}_{-0.72}^{+1.05}$& 4.0\\
NL1& ${3.52}_{-0.59}^{+0.83}$ & 0.5\\
NL2& ${2.45}_{-0.32}^{+0.42}$& 0.05\\
NL3& ${1.64}_{-0.17}^{+0.21}$&  0.2\\
SL3& ${1.33}_{-0.15}^{+0.18}$&  0.01\\
SL2&${1.94}_{-0.23}^{+0.29}$&    0.03\\
SL1& ${2.40}_{-0.32}^{+0.42}$& 1.6\\
SL0&${3.52}_{-0.60}^{+0.85}$&3.2\\
S-HotSpot& ${5.43}_{-1.21}^{+ 2.00}$& 0.1\\
\hline
\noalign{\smallskip}
\end{tabular}
\end{center}
\end{table}

\begin{figure*}[ht!]
\begin{center}
\includegraphics[width=14cm]{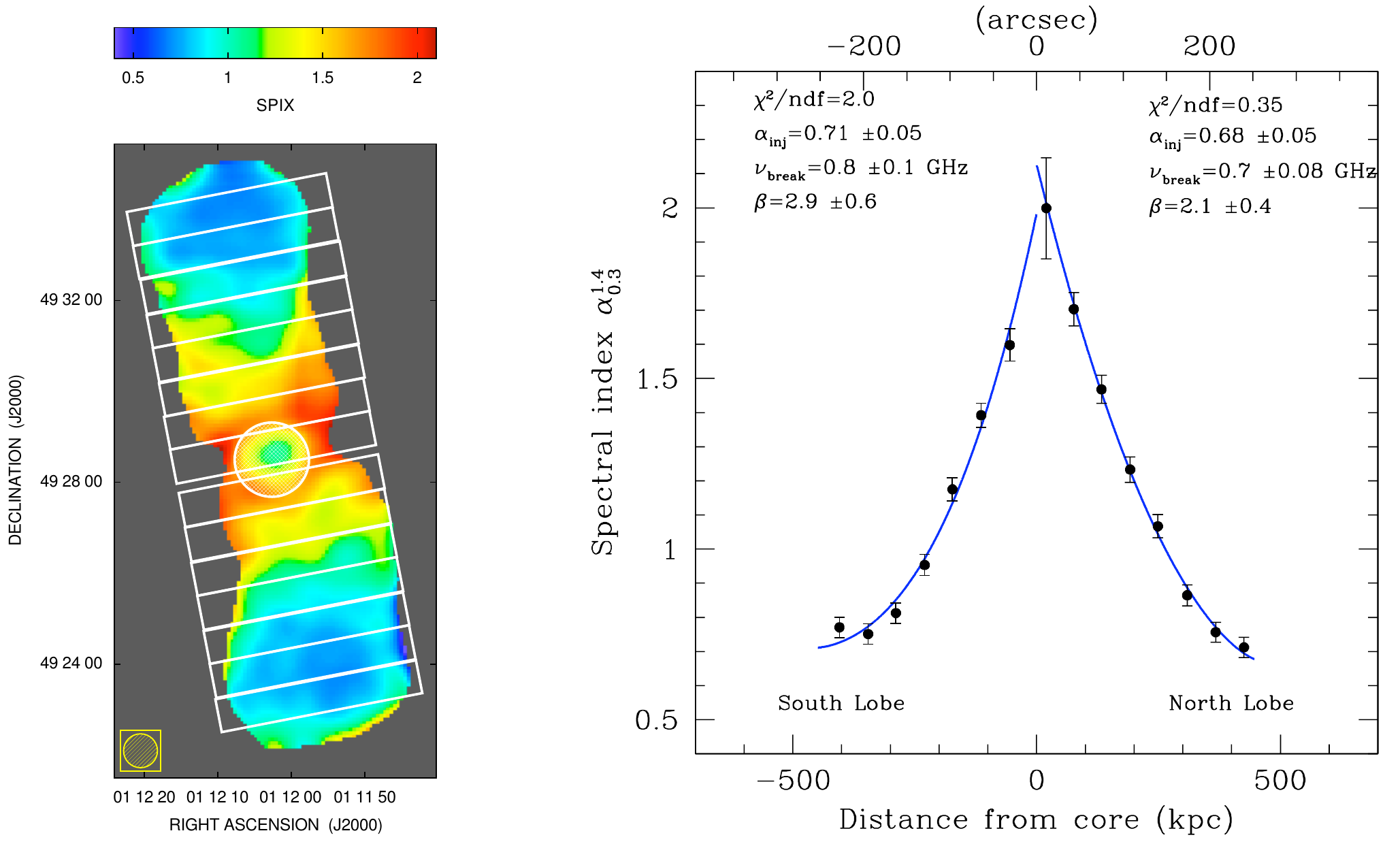}
\end{center}
\caption[]{3C35:  {\it left} in color the spectral index map between 327\,Mhz and 1.4\,GHz; 
overlayed the white array of boxes where the measures have been
taken. 
{\it Right} fit of the spectral index $\alpha_{1.4}^{0.3}$ with the
synchrotron model.}
\label{3c35spix_fit}
\end{figure*} 

\begin{figure*}[ht!]
\begin{center}
\includegraphics[width=14cm]{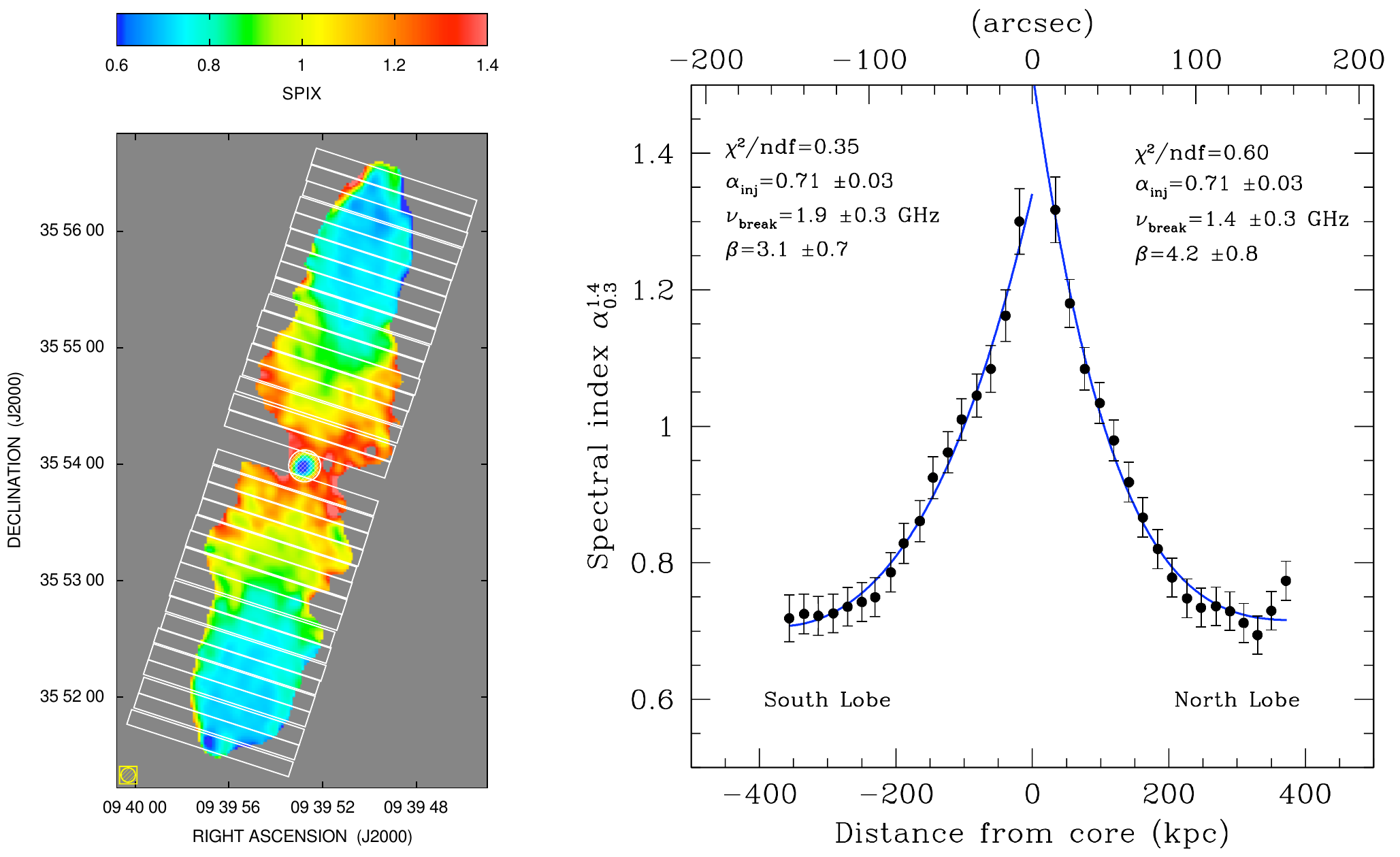}
\end{center}
\caption[]{3C223:  {\it left} in color the spectral index map between 327\,Mhz and 1.48\,GHz; 
overlayed the white array of boxes where the measures have been
taken. {\it Right} fit of the spectral index $\alpha_{1.4}^{0.3}$ with the
  synchrotron model.}
\label{3c223spix_fit}
\end{figure*} 

\subsection{Fit of the spectral index profile}
In this section we evaluate, independently for the two lobes of each
source, the frequency break and the injection spectral index, by
fitting with the synchrotron model the two frequency spectral index,
$\alpha_{1.4}^{0.3}$, as a function of the distance from the core.
This method allows an investigation of the spectral index behaviour at
higher spatial resolution then we used in the previously described,
multi-frequency analysis (Sect. \ref{multifreq}), which was limited by
the resolution of the 74\,MHz images.  Following the assumption that
the sources are expanding with a constant velocity, we have that
$\nu_{break} \propto 1/t^2 \propto 1/d^2$ since t=d/v. Here t is the
age and d is the distance from the hot-spot. We fitted $\nu_{break}
\propto 1/d^{\beta}$, where $\beta$ is a free parameter. If $\beta$ is
2, this means that the source is expanding with a constant velocity
and adiabatic losses are negligible.  If $\beta$ is steeper than 2
either the expansion losses play an important role in the energetic of
the source and/or the radio source is not expanding with a constant
speed.

The fit for the source 3C35 (Fig.\ref{3c35spix_fit}) gives
$\nu_{break}\,\simeq$\,0.7-0.8\,GHz which is in good agreement with
the one found in Sect.\,4.1. $\alpha_{inj}$\,=\,0.7 is slightly
steeper with respect to the $\alpha_{inj}$ found in Sect.\,4.1, but
still consistent within the errors. For this source the value of the
parameter $\beta$ is close to a value of about 2.5$\pm$0.7 this means that
adiabatic losses are negligible for 3C35, this can be further
confirmed by the tubular structure of the source.

For the source 3C223 (Fig.\ref{3c223spix_fit}) we found
$\nu_{break}\simeq$\,1.4 -1.9\,GHz, which is consistent with the one
found fitting the spectral shape in Sect.\,4.1. In this case the
injection index of 0.71 is steeper than the assumed
$\alpha_{inj}$\,=0.5 used previously (Sect.\,4.1) and probably
reflects both the higher frequency range of this method and the
non-power-law shape of the hot-spots in this source.  The parameter
$\beta$ for this source is about 3.6$\pm$1.1. As noted above, $\beta>$ 2
indicates either expansion energy losses or a source expanding at a
non-constant speed. A physical motivation for such a value could be
explained as follows. If the inverse Compton dominates the radiative
losses, $\epsilon_{break}\,\propto$\,1/t while $\nu_{break}\,\propto$
B$\cdot\epsilon_{break}^2$, where $\epsilon_{break}$ is the energy
break and B is the magnetic field. The magnetic field decreases with
the lateral expansion of the lobes as B\,$\propto$\,1/$R^2$, where R
is the radius of section of the lobe. Since the source evolves in a
self-similar way \cite[]{Kaiser&Alexander1997}, R is proportional to
d, which gives B\,$\propto$\,1/$d^2$, and therefore
$\nu_{break}\,\propto$1/$d^4$.  That is the value that we found for
the free parameter $\beta$. In this case the action of expansion
losses can be seen in the arrow structure of the source.

\section{Summary and conclusions}

In this paper we present new VLA images of the sources 3C35 and 3C223
at the observing frequencies of 327 and 74\,MHz.

By combining our images with those at 1.4\,GHz available in the
literature, we produced spectral index distribution maps between
74--327\,MHz and 327\,MHz--1.4\,GHz for both sources. The spectral
index across the sources are more constant in the low frequency range,
while in the high frequency range the spectral indices increase from
the hot-spots to the inner region of the lobes near to the core. In
particular, for the source 3C35, $\alpha$ ranges between 0.6 and 0.8
in the interval of frequencies 74--327\,MHz, while between the
frequencies 327\,MHz--1.4\,GHz the values of $\alpha$ change from 0.6
in the hot-spot's region to 1.7 in the inner region of the lobes. On
the other hand, for the source 3C223 the value of $\alpha$ is on
average 0.6 in the range 74--327\,MHz , but it could reach extreme
values which range between 0.4 and 1.6.  In the range between
327\,MHz--1.4\,GHz $\alpha$ varies from 0.7 in the hot-spots to 1.5 in
the inner region of the lobes.

By considering the two radio sources in a minimum energy condition,
i.e. in the equipartition regime, we estimated the magnetic field of
the two sources.  The estimate was made using two different
approaches often adopted in the literature, a fixed frequency range
and a fixed energy range. Moreover, two different
plasma populations has been considered (see Tab. \ref{35flux} and \ref{223flux}): one in
which the energy is equally divided between relativistic protons and
electrons and another one in which all the energy is provided by a
plasma of relativistic electrons--positrons.  For both sources the
resulting equipartition magnetic field ranges between values of
0.5--1.6 $\mu$G, in concordance with typical values of the measured IC
magnetic fields. In particular, in the case of 3C223, the value of the
equipartition magnetic field is within a factor of two in agreement with the measured IC magnetic
field \cite[]{Croston2004}.  

By using our images with those at higher frequencies available in the
literature, we obtained the spectral shape of the radio spectrum in
many different positions along the lobes. The hot spots of the source
3C35 are well described by power--laws, while the hot spots of 3C223
show quite curved spectra. The inner regions of the lobes for the two
sources present a break in the range of frequency around 1.0\,GHz.
 
Since for both the sources the magnetic field is low, the inverse
Compton losses are as important as the synchrotron losses, and we can
assume an isotropic electron population. Therefore we fitted the
spectra with a JP \cite[]{JP} model to estimate the frequency break
$\nu_{break}$; in the case of 3C35 we also estimated $\alpha_{inj}$
while for 3C223 we used a fixed value. In the case of 3C35 we found
that $\nu_{break} \simeq$ 800\,MHz and $\alpha_{inj}$ is on average
0.5. For 3C223 the $\nu_{break}$ is about 1.4\,GHz with a fixed
$\alpha_{inj}$=0.5.

The break frequency $\nu_{break}$ is a time-dependent function, by
assuming that there is no expansion and the magnetic field is constant,
from the frequency break we calculated the radiative age
of the source $t_{syn}$.  \cite{Blundell2000} claimed an anomalous
diffusion of relativistic particles which implies that no information
about the age of the source can be inferred from the shape of the
emission radio spectrum. They discussed about the discrepancy between
the estimates of the spectral and dynamical ages for sources older
then $10^{7}$ yr. But \cite{Kaiser2000} demonstrated that diffusion
will not alter the distribution of relativistic particles, therefore
the spatial distribution of the synchrotron radio emission can be use
to estimate the age for FRII sources \cite[]{Fanaroff1974}.  Moreover,
as we discussed above, the magnetic field of these particular sources
is low with respect to the inverse Compton equivalent magnetic field,
therefore a spatially variable magnetic field has a minor impact on
the energy losses of the relativistic electrons. 
For 3C35 the estimated age is about 143$\pm$20\,Myr while for 3C223 is about
 72$\pm$4\,Myr.  The radiative age confirms that the two sources are rather
old. However, these estimates must be considered upper limits if
adiabatic losses can not be neglected. 

A high resolution analysis of the spectral index behaviour has been
made by fitting the two frequency spectral index $\alpha_{1.4}^{0.3}$
with the synchrotron model as function of the distance from the
core. We fitted with a law $\nu_{break} \propto 1/d^{\beta}$. For the
source 3C35 the frequency break is about 800\,MHz and parameter
$\beta$ is about 2.5$\pm$0.7, in agreement with an expansion with a
constant speed and/or unimportant adiabatic losses.  For the source
3C223, $\nu_{break} \simeq$ 1.4\,GHz, while the parameter $\beta$ is
about 3.6$\pm$1.1; as discussed before this could be explained if adiabatic
losses play an important role in the energy balance of the source
and/or if the expansion velocity of the source is not constant.

\begin{acknowledgements}
We would like to thank the anonymous referee for insightful comments
which improved this manuscript.
E.O.acknowledges financial support of Austrian Science Foundation
(FWF) through grant number P18523-N16.
The National Radio Astronomy Observatory is
operated by Associated Universities, Inc.,
under contract to the National Science
 Fundation.
This research has made use of the NASA/IPAC Extragalactic Database (NED) which 
operated by the Jet Propulsion Laboratory, California Institute of Technology,
under contract with the National Aeronautics and Space Administration
and of CATS database Astrophysical CATalogs support System. 
\end{acknowledgements}

\bibliography{pap_orru.bbl}

\begin{thebibliography}{58}
\expandafter\ifx\csname natexlab\endcsname\relax\def\natexlab#1{#1}\fi

\bibitem[{{Abazajian} {et~al.}(2009){Abazajian}, {Adelman-McCarthy},
  {Ag{\"u}eros}, {Allam}, {Allende Prieto}, {An}, {Anderson}, {Anderson},
  {Annis}, {Bahcall}, {Bailer-Jones}, {Barentine}, {Bassett}, {Becker},
  {Beers}, {Bell}, {Belokurov}, {Berlind}, {Berman}, {Bernardi}, {Bickerton},
  {Bizyaev}, {Blakeslee}, {Blanton}, {Bochanski}, {Boroski}, {Brewington},
  {Brinchmann}, {Brinkmann}, {Brunner}, {Budav{\'a}ri}, {Carey}, {Carliles},
  {Carr}, {Castander}, {Cinabro}, {Connolly}, {Csabai}, {Cunha}, {Czarapata},
  {Davenport}, {de Haas}, {Dilday}, {Doi}, {Eisenstein}, {Evans}, {Evans},
  {Fan}, {Friedman}, {Frieman}, {Fukugita}, {G{\"a}nsicke}, {Gates},
  {Gillespie}, {Gilmore}, {Gonzalez}, {Gonzalez}, {Grebel}, {Gunn},
  {Gy{\"o}ry}, {Hall}, {Harding}, {Harris}, {Harvanek}, {Hawley}, {Hayes},
  {Heckman}, {Hendry}, {Hennessy}, {Hindsley}, {Hoblitt}, {Hogan}, {Hogg},
  {Holtzman}, {Hyde}, {Ichikawa}, {Ichikawa}, {Im}, {Ivezi{\'c}}, {Jester},
  {Jiang}, {Johnson}, {Jorgensen}, {Juri{\'c}}, {Kent}, {Kessler}, {Kleinman},
  {Knapp}, {Konishi}, {Kron}, {Krzesinski}, {Kuropatkin}, {Lampeitl},
  {Lebedeva}, {Lee}, {Lee}, {Leger}, {L{\'e}pine}, {Li}, {Lima}, {Lin}, {Long},
  {Loomis}, {Loveday}, {Lupton}, {Magnier}, {Malanushenko}, {Malanushenko},
  {Mandelbaum}, {Margon}, {Marriner}, {Mart{\'{\i}}nez-Delgado}, {Matsubara},
  {McGehee}, {McKay}, {Meiksin}, {Morrison}, {Mullally}, {Munn}, {Murphy},
  {Nash}, {Nebot}, {Neilsen}, {Newberg}, {Newman}, {Nichol}, {Nicinski},
  {Nieto-Santisteban}, {Nitta}, {Okamura}, {Oravetz}, {Ostriker}, {Owen},
  {Padmanabhan}, {Pan}, {Park}, {Pauls}, {Peoples}, {Percival}, {Pier}, {Pope},
  {Pourbaix}, {Price}, {Purger}, {Quinn}, {Raddick}, {Fiorentin}, {Richards},
  {Richmond}, {Riess}, {Rix}, {Rockosi}, {Sako}, {Schlegel}, {Schneider},
  {Scholz}, {Schreiber}, {Schwope}, {Seljak}, {Sesar}, {Sheldon}, {Shimasaku},
  {Sibley}, {Simmons}, {Sivarani}, {Smith}, {Smith}, {Smol{\v c}i{\'c}},
  {Snedden}, {Stebbins}, {Steinmetz}, {Stoughton}, {Strauss}, {Subba Rao},
  {Suto}, {Szalay}, {Szapudi}, {Szkody}, {Tanaka}, {Tegmark}, {Teodoro},
  {Thakar}, {Tremonti}, {Tucker}, {Uomoto}, {Vanden Berk}, {Vandenberg},
  {Vidrih}, {Vogeley}, {Voges}, {Vogt}, {Wadadekar}, {Watters}, {Weinberg},
  {West}, {White}, {Wilhite}, {Wonders}, {Yanny}, {Yocum}, {York}, {Zehavi},
  {Zibetti}, \& {Zucker}}]{SDSS}
{Abazajian}, K.~N., {Adelman-McCarthy}, J.~K., {Ag{\"u}eros}, M.~A., {et~al.}
  2009, \apjs, 182, 543

\bibitem[{{Baum} {et~al.}(1988){Baum}, {Heckman}, {Bridle}, {van Breugel}, \&
  {Miley}}]{Baum1988}
{Baum}, S.~A., {Heckman}, T.~M., {Bridle}, A., {van Breugel}, W.~J.~M., \&
  {Miley}, G.~K. 1988, \apjs, 68, 643

\bibitem[{{Beck} \& {Krause}(2005)}]{Beck2005}
{Beck}, R. \& {Krause}, M. 2005, Astronomische Nachrichten, 326, 414

\bibitem[{{Begelman} {et~al.}(1984){Begelman}, {Blandford}, \&
  {Rees}}]{Begelman1984}
{Begelman}, M.~C., {Blandford}, R.~D., \& {Rees}, M.~J. 1984, Reviews of Modern
  Physics, 56, 255

\bibitem[{{Blandford} \& {Rees}(1974)}]{Blandford&Rees1974}
{Blandford}, R.~D. \& {Rees}, M.~J. 1974, \mnras, 169, 395

\bibitem[{{Blundell} \& {Rawlings}(2000)}]{Blundell2000}
{Blundell}, K.~M. \& {Rawlings}, S. 2000, \aj, 119, 1111

\bibitem[{{Blundell} {et~al.}(1999){Blundell}, {Rawlings}, \&
  {Willott}}]{Blundell1999}
{Blundell}, K.~M., {Rawlings}, S., \& {Willott}, C.~J. 1999, \aj, 117, 677

\bibitem[{{Brunetti} {et~al.}(1997){Brunetti}, {Setti}, \&
  {Comastri}}]{Brunetti1997}
{Brunetti}, G., {Setti}, G., \& {Comastri}, A. 1997, \aap, 325, 898

\bibitem[{{Burbidge} \& {Strittmatter}(1972)}]{Burbidge1972}
{Burbidge}, E.~M. \& {Strittmatter}, P.~A. 1972, \apjl, 172, L37+

\bibitem[{{Carilli} {et~al.}(1991){Carilli}, {Perley}, {Dreher}, \&
  {Leahy}}]{Carilli1991}
{Carilli}, C.~L., {Perley}, R.~A., {Dreher}, J.~W., \& {Leahy}, J.~P. 1991,
  \apj, 383, 554

\bibitem[{{Cohen} {et~al.}(2007){Cohen}, {Lane}, {Cotton}, {Kassim}, {Lazio},
  {Perley}, {Condon}, \& {Erickson}}]{Aaron2007}
{Cohen}, A.~S., {Lane}, W.~M., {Cotton}, W.~D., {et~al.} 2007, \aj, 134, 1245

\bibitem[{{Condon} {et~al.}(1998){Condon}, {Cotton}, {Greisen}, {Yin},
  {Perley}, {Taylor}, \& {Broderick}}]{NVSS}
{Condon}, J.~J., {Cotton}, W.~D., {Greisen}, E.~W., {et~al.} 1998, \aj, 115,
  1693

\bibitem[{{Cornwell} \& {Perley}(1992)}]{Cornwell1992}
{Cornwell}, T.~J. \& {Perley}, R.~A. 1992, \aap, 261, 353

\bibitem[{{Cotton} {et~al.}(2004){Cotton}, {Condon}, {Perley}, {Kassim},
  {Lazio}, {Cohen}, {Lane}, \& {Erickson}}]{Cotton2004}
{Cotton}, W.~D., {Condon}, J.~J., {Perley}, R.~A., {et~al.} 2004, in
  Ground-based Telescopes. Edited by Oschmann, Jacobus M., Jr. Proceedings of
  the SPIE, Volume 5489, pp. 180-189 (2004)., ed. J.~M. {Oschmann}, Jr.,
  180--189

\bibitem[{{Croston} {et~al.}(2004){Croston}, {Birkinshaw}, {Hardcastle}, \&
  {Worrall}}]{Croston2004}
{Croston}, J.~H., {Birkinshaw}, M., {Hardcastle}, M.~J., \& {Worrall}, D.~M.
  2004, \mnras, 353, 879

\bibitem[{{Croston} {et~al.}(2005){Croston}, {Hardcastle}, {Harris}, {Belsole},
  {Birkinshaw}, \& {Worrall}}]{Croston2005}
{Croston}, J.~H., {Hardcastle}, M.~J., {Harris}, D.~E., {et~al.} 2005, \apj,
  626, 733

\bibitem[{{Eilek} {et~al.}(1997){Eilek}, {Melrose}, \& {Walker}}]{Eilek1997}
{Eilek}, J.~A., {Melrose}, D.~B., \& {Walker}, M.~A. 1997, \apj, 483, 282

\bibitem[{{Fanaroff} \& {Riley}(1974)}]{Fanaroff1974}
{Fanaroff}, B.~L. \& {Riley}, J.~M. 1974, \mnras, 167, 31P

\bibitem[{{Goodger} {et~al.}(2008){Goodger}, {Hardcastle}, {Croston}, {Kassim},
  \& {Perley}}]{Goodger2008}
{Goodger}, J.~L., {Hardcastle}, M.~J., {Croston}, J.~H., {Kassim}, N.~E., \&
  {Perley}, R.~A. 2008, \mnras, 386, 337

\bibitem[{{Hardcastle} \& {Croston}(2005)}]{Hardcastle2005}
{Hardcastle}, M.~J. \& {Croston}, J.~H. 2005, \mnras, 363, 649

\bibitem[{{Hogbom}(1979)}]{Hogbom1979}
{Hogbom}, J.~A. 1979, \aaps, 36, 173

\bibitem[{{Ishwara-Chandra} \& {Saikia}(1999)}]{Ishwara-Chandra1999}
{Ishwara-Chandra}, C.~H. \& {Saikia}, D.~J. 1999, \mnras, 309, 100

\bibitem[{{Isobe} {et~al.}(2002){Isobe}, {Tashiro}, {Makishima}, {Iyomoto},
  {Suzuki}, {Murakami}, {Mori}, \& {Abe}}]{Isobe2002}
{Isobe}, N., {Tashiro}, M., {Makishima}, K., {et~al.} 2002, \apjl, 580, L111

\bibitem[{{Jaffe} \& {Perola}(1973)}]{JP}
{Jaffe}, W.~J. \& {Perola}, G.~C. 1973, \aap, 26, 423

\bibitem[{{J\"{a}gers}(1987)}]{Jaegers1987}
{J\"{a}gers}, W.~J. 1987, \aaps, 67, 395

\bibitem[{{Jamrozy} {et~al.}(2004){Jamrozy}, {Klein}, {Mack}, {Gregorini}, \&
  {Parma}}]{Jamrozy2004}
{Jamrozy}, M., {Klein}, U., {Mack}, K.-H., {Gregorini}, L., \& {Parma}, P.
  2004, \aap, 427, 79

\bibitem[{{Jamrozy} {et~al.}(2008){Jamrozy}, {Konar}, {Machalski}, \&
  {Saikia}}]{Jamrozy2008}
{Jamrozy}, M., {Konar}, C., {Machalski}, J., \& {Saikia}, D.~J. 2008, \mnras,
  385, 1286

\bibitem[{{Jamrozy} {et~al.}(2005){Jamrozy}, {Machalski}, {Mack}, \&
  {Klein}}]{Jamrozy2005}
{Jamrozy}, M., {Machalski}, J., {Mack}, K.-H., \& {Klein}, U. 2005, \aap, 433,
  467

\bibitem[{{Kaiser}(2000)}]{Kaiser2000}
{Kaiser}, C.~R. 2000, \aap, 362, 447

\bibitem[{{Kaiser} \& {Alexander}(1997)}]{Kaiser&Alexander1997}
{Kaiser}, C.~R. \& {Alexander}, P. 1997, \mnras, 286, 215

\bibitem[{{Kaiser} {et~al.}(1997){Kaiser}, {Dennett-Thorpe}, \&
  {Alexander}}]{KaiserThorpe1997}
{Kaiser}, C.~R., {Dennett-Thorpe}, J., \& {Alexander}, P. 1997, \mnras, 292,
  723

\bibitem[{{Kardashev}(1962)}]{K}
{Kardashev}, N.~S. 1962, Soviet Astronomy, 6, 317

\bibitem[{{Kassim} {et~al.}(2007){Kassim}, {Lazio}, {Erickson}, {Perley},
  {Cotton}, {Greisen}, {Cohen}, {Hicks}, {Schmitt}, \& {Katz}}]{Namir2007}
{Kassim}, N.~E., {Lazio}, T.~J.~W., {Erickson}, W.~C., {et~al.} 2007, \apjs,
  172, 686

\bibitem[{{Kassim} {et~al.}(1993){Kassim}, {Perley}, {Erickson}, \&
  {Dwarakanath}}]{kassim1993}
{Kassim}, N.~E., {Perley}, R.~A., {Erickson}, W.~C., \& {Dwarakanath}, K.~S.
  1993, \aj, 106, 2218

\bibitem[{{Kellermann}(1964)}]{Kellermann1964}
{Kellermann}, K.~I. 1964, \apj, 140, 969

\bibitem[{{Konar} {et~al.}(2009){Konar}, {Hardcastle}, {Croston}, \&
  {Saikia}}]{Konar2009}
{Konar}, C., {Hardcastle}, M.~J., {Croston}, J.~H., \& {Saikia}, D.~J. 2009,
  \mnras, 400, 480

\bibitem[{{Laing} \& {Peacock}(1980)}]{Laing1980}
{Laing}, R.~A. \& {Peacock}, J.~A. 1980, \mnras, 190, 903

\bibitem[{{Laing} {et~al.}(1983){Laing}, {Riley}, \& {Longair}}]{Laing1983}
{Laing}, R.~A., {Riley}, J.~M., \& {Longair}, M.~S. 1983, \mnras, 204, 151

\bibitem[{{Lane} {et~al.}(2005){Lane}, {Cohen}, {Kassim}, {Lazio}, {Perley},
  {Cotton}, \& {Greisen}}]{Lane2005}
{Lane}, W.~M., {Cohen}, A.~S., {Kassim}, N.~E., {et~al.} 2005, Radio Science,
  40, 5

\bibitem[{{Lara} {et~al.}(2004){Lara}, {Giovannini}, {Cotton}, {Feretti},
  {Marcaide}, {M{\'a}rquez}, \& {Venturi}}]{Lara2004}
{Lara}, L., {Giovannini}, G., {Cotton}, W.~D., {et~al.} 2004, \aap, 421, 899

\bibitem[{{Leahy} \& {Perley}(1991)}]{Leahy1991}
{Leahy}, J.~P. \& {Perley}, R.~A. 1991, \aj, 102, 537

\bibitem[{{Machalski} {et~al.}(2007){Machalski}, {Chy{\.z}y}, {Stawarz}, \&
  {Kozie{\l}}}]{Machalski2007}
{Machalski}, J., {Chy{\.z}y}, K.~T., {Stawarz}, {\L}., \& {Kozie{\l}}, D. 2007,
  \aap, 462, 43

\bibitem[{{Mason} {et~al.}(2000){Mason}, {Carrera}, {Hasinger}, {Andernach},
  {Aragon-Salamanca}, {Barcons}, {Bower}, {Brandt}, {Branduardi-Raymont},
  {Burgos-Mart{\'{\i}}n}, {Cabrera-Guerra}, {Carballo}, {Castander}, {Ellis},
  {Gonz{\'a}lez-Serrano}, {Mart{\'{\i}}nez-Gonz{\'a}lez},
  {Mart{\'{\i}}n-Mirones}, {McMahon}, {Mittaz}, {Nicholson}, {Page},
  {P{\'e}rez-Fournon}, {Puchnarewicz}, {Romero-Colmenero}, {Schwope}, {Vila},
  {Watson}, \& {Wonnacott}}]{Mason2000}
{Mason}, K.~O., {Carrera}, F.~J., {Hasinger}, G., {et~al.} 2000, \mnras, 311,
  456

\bibitem[{{Meli} {et~al.}(2008){Meli}, {Becker}, \& {Quenby}}]{Meli2008}
{Meli}, A., {Becker}, J.~K., \& {Quenby}, J.~J. 2008, \aap, 492, 323

\bibitem[{{Murgia}(2000)}]{MurgiaPhD}
{Murgia}, M., 2000, PhD Thesis, University of Bologna

\bibitem[{{Pacholczyk}(1970)}]{P}
{Pacholczyk}, A.~G. 1970, {Radio astrophysics. Nonthermal processes in galactic
  and extragalactic sources} (Series of Books in Astronomy and Astrophysics,
  San Francisco: Freeman, 1970)

\bibitem[{{Parma} {et~al.}(1999){Parma}, {Murgia}, {Morganti}, {Capetti}, {de
  Ruiter}, \& {Fanti}}]{Parma1999}
{Parma}, P., {Murgia}, M., {Morganti}, R., {et~al.} 1999, \aap, 344, 7

\bibitem[{{Perley}(1999)}]{Perley1999}
{Perley}, R.~A. 1999, in ASP Conf. Ser. 180: Synthesis Imaging in Radio
  Astronomy II, ed. G.~B. {Taylor}, C.~L. {Carilli}, \& R.~A. {Perley}, 383

\bibitem[{{Rengelink} {et~al.}(1997){Rengelink}, {Tang}, {de Bruyn}, {Miley},
  {Bremer}, {Roettgering}, \& {Bremer}}]{WENSS}
{Rengelink}, R.~B., {Tang}, Y., {de Bruyn}, A.~G., {et~al.} 1997, \aaps, 124,
  259

\bibitem[{{Rudnick} {et~al.}(1994){Rudnick}, {Katz-Stone}, \&
  {Anderson}}]{Rudnick1994}
{Rudnick}, L., {Katz-Stone}, D.~M., \& {Anderson}, M.~C. 1994, \apjs, 90, 955

\bibitem[{{Saripalli} {et~al.}(2005){Saripalli}, {Hunstead}, {Subrahmanyan}, \&
  {Boyce}}]{Saripalli2005}
{Saripalli}, L., {Hunstead}, R.~W., {Subrahmanyan}, R., \& {Boyce}, E. 2005,
  \aj, 130, 896

\bibitem[{{Scheuer}(1974)}]{Scheuer1974}
{Scheuer}, P.~A.~G. 1974, \mnras, 166, 513

\bibitem[{{Schoenmakers} {et~al.}(2001){Schoenmakers}, {de Bruyn},
  {R{\"o}ttgering}, \& {van der Laan}}]{Schoenmakers2001}
{Schoenmakers}, A.~P., {de Bruyn}, A.~G., {R{\"o}ttgering}, H.~J.~A., \& {van
  der Laan}, H. 2001, \aap, 374, 861

\bibitem[{{Schoenmakers} {et~al.}(2000){Schoenmakers}, {Mack}, {de Bruyn},
  {R{\"o}ttgering}, {Klein}, \& {van der Laan}}]{Schoenmakers2000}
{Schoenmakers}, A.~P., {Mack}, K.-H., {de Bruyn}, A.~G., {et~al.} 2000, \aaps,
  146, 293

\bibitem[{{Slee} {et~al.}(2001){Slee}, {Roy}, {Murgia}, {Andernach}, \&
  {Ehle}}]{Slee2001}
{Slee}, O.~B., {Roy}, A.~L., {Murgia}, M., {Andernach}, H., \& {Ehle}, M. 2001,
  \aj, 122, 1172

\bibitem[{{Spergel} {et~al.}(2003){Spergel}, {Verde}, {Peiris}, {Komatsu},
  {Nolta}, {Bennett}, {Halpern}, {Hinshaw}, {Jarosik}, {Kogut}, {Limon},
  {Meyer}, {Page}, {Tucker}, {Weiland}, {Wollack}, \& {Wright}}]{Spergel2003}
{Spergel}, D.~N., {Verde}, L., {Peiris}, H.~V., {et~al.} 2003, \apjs, 148, 175

\bibitem[{{Tribble}(1993)}]{Tribble1993}
{Tribble}, P.~C. 1993, \mnras, 261, 57

\bibitem[{{van Breugel} \& {J\"{a}gers}(1982)}]{vanBreugel1982}
{van Breugel}, W. \& {J\"{a}gers}, W. 1982, \aaps, 49, 529

\end{thebibliography}
\bibliographystyle{aa} 

\appendix
\section{Faint radio galaxy}
\begin{figure}[h]
\begin{center}
\includegraphics[width=8cm]{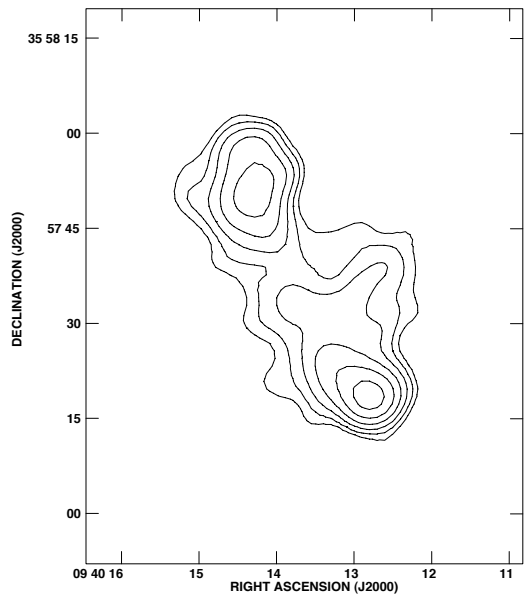}
\end{center}
\caption[]{327 MHz VLA image. The resolution is
  7\arcsec$\times$6\arcsec\, with a PA=$-78$\degr, contours start at
  (3$\sigma$) and are scaled by $\sqrt{2}$, the first level of contours is 1.8\,mJy/beam.}
\label{QSOcntr}
\end{figure}

 In the field of 3C223 at 327\,MHz, at the position RA 09$^h$ 40$^m$
 13.9$^s$ Dec +35\degr\ 57\arcmin\ 34\arcsec, we detected a small
 double-lobed radio source. This radio source has already been
 detected (see Fig.\,6 in \cite{Croston2004}) in the 1.4\,GHz VLA data
 by \cite{Leahy1991}. In the same position, a bright X-ray source has
 been found in XMM-{\it Newton} by \cite{Croston2004}. This X-ray
 source, previously detected with {\it ROSAT}, was at first
 erroneously identified with a star which is offset by 20 arcsec from
 the X-ray source \cite[]{Mason2000}. A faint optical object
 coincident with the X-ray source has been detected in the Sloan
 Digital Sky Survey (the u-band magnitude is $\simeq$\,22).
 Fig.\,\ref{QSOcntr} shows the 327\,MHz image of the faint radio
 source with resolution of 7\arcsec$\times$6\arcsec. This image points
 out an extended low brightness emission between the two lobes in the
 region around the west side of the core. This implies the presence of
 low energy electrons in this region, which possibly might contribute
 to the IC emission. This emission was weakly detected in the image at
 1.4\,GHz. The total flux of the radio source at 327\,MHz is
 $\simeq$\,93\,$\pm$\,2.5\,mJy. The angular extension is about one
 arcmin. Since there is no redshift measured for the optical
 counterpart, we could not provide an estimate of the linear
 dimensions or the equipartition magnetic field.  The detection and
 the study of the low brightness emission of these kinds of sources,
 which show most of their emission in the low frequency range, is one
 of the main goals of the new generation of radio telescopes like
 LOFAR, SKA, LWA etc.
 
\end{document}